\documentclass[aps, prd, 10pt, twocolumn, superscriptaddress, noshowpacs, preprintnumbers, longbibliography, groupedaddress, footinbib, bibnotes]{revtex4-1}

\usepackage{amsmath}
\usepackage{amsfonts}
\usepackage{amssymb}
\usepackage{bbold}
\usepackage{epsfig}
\usepackage{graphicx}
\usepackage{bm}
\usepackage{array}
\usepackage{hyperref}
\usepackage{listings}
\usepackage{float}
\usepackage[normalem]{ulem}
\usepackage[latin1]{inputenc}
\usepackage{dsfont} % gives real R and complex C.
\usepackage[english]{babel}
\usepackage[normalem]{ulem}
\usepackage{color}

\newcommand{\eref}[1]{Eq.~(\ref{#1})}
\newcommand{\fref}[1]{Fig.~\ref{#1}}
\newcommand{\sref}[1]{Sec.~\ref{#1}}
\newcommand{\tref}[1]{Tab.~\ref{#1}}

\begin{document}

\title{Enhancement or damping of fast neutrino flavor conversions due to collisions}

\author{Rasmus S.~L.~Hansen}%
\email{rslhansen@nbi.ku.dk}
\author{Shashank Shalgar}%
\email{shashank.shalgar@nbi.ku.dk}
\author{Irene Tamborra}%
\email{tamborra@nbi.ku.dk}
\affiliation{%
 Niels Bohr International Academy and DARK, Niels Bohr Institute, University of Copenhagen, Blegdamsvej 17, 2100, Copenhagen, Denmark
}%

\begin{abstract}
Fast neutrino flavor conversion can occur in core-collapse supernovae or compact binary merger remnants when non-forward collisions are also at play, and neutrinos are not fully decoupled from matter. 
This work aims to shed light on the conditions under which fast flavor conversion is enhanced or suppressed by collisions. By relying on a neutrino toy model with three angular bins in the absence of spatial inhomogeneities, we consider two  angular configurations: The first one with angular distributions of $\nu_e$ and $\bar\nu_e$ that are almost isotropic as expected before complete neutrino decoupling and showing little flavor conversion when collisions are absent. The second one with angular distributions of $\nu_e$ and $\bar\nu_e$ that are forward peaked as expected in the free-streaming regime and showing significant flavor conversion in the absence of collisions. By including angle-independent, direction-changing collisions, we find that collisions are responsible for an overall enhancement (damping) of flavor conversion in the former (latter) angular configuration. These opposite outcomes are due to the non-trivial interplay between collisions, flavor conversion, and the initial angular distributions of the electron type neutrinos. 
The enhancement in neutrino flavor conversion is found to be anticorrelated with the magnitude of flavor conversions in the absence of collisions.

\end{abstract}

\maketitle

%%%%%%%%%%%%%%%%%%%%%%%%%%%%%%%%%%%%%%%%%%%%%%%%%%%%%%%%%%%%%%%%%%%%%%%%%%%%%%%%%%%%%%%%%%%%%%%%
\section{Introduction}

The physics linked to neutrino flavor evolution in dense astrophysical environments remains full of mysteries despite intense theoretical work~\cite{Tamborra:2020cul,Capozzi:2022slf,Mirizzi:2015eza,Duan:2010bg}. 
One of the main complications is related to the modeling of neutrino flavor evolution in the presence of  coherent forward scattering of neutrinos among themselves. 
In addition, the flavor evolution history may be affected by collisions with the background medium~\cite{Stodolsky:1986dx,Stodolsky:1974hm}, possibly leading to  loss of coherence in the flavor evolution when the mean-free-path of neutrinos is much smaller than the typical length scale associated with neutrino flavor evolution.

In  core-collapse supernovae, neutrino flavor transformation among the  active flavors was first expected to  occur at large distances from the proto-neutron star, where neutrinos are in the free streaming regime~\cite{Mirizzi:2015eza,Duan:2010bg}. In such a scenario,   collisions with the background medium were considered to be negligible, except for the occasional direction-changing scatterings  with the matter envelope, leading to the formation of a ``neutrino halo''~\cite{Cherry:2012zw}. The latter results in a broadening of the neutrino angular distribution, potentially  affecting  $\nu$--$\nu$ refraction at distances larger than $\mathcal{O}(100)$~km from the supernova core. Reference~\cite{Sarikas:2012vb} reported multi-angle matter suppression of flavor conversion due to the neutrino halo by employing the linear stability analysis, while other attempts~\cite{Cherry:2013mv,Cirigliano:2018rst,Zaizen:2019ufj,Cherry:2019vkv} to model the impact of the neutrino halo on flavor conversion involve various approximations. Hence, a robust assessment on the relevance of the neutrino halo for flavor conversion is still lacking. 

The interplay between collisions and flavor conversion has been revisited recently in the context of fast flavor conversions triggered by the pairwise scattering of neutrinos at high densities, deep in the core of compact objects~\cite{Tamborra:2020cul,Chakraborty:2016yeg}. Fast flavor conversions, as the name suggests, develop over characteristic time scales that are very small as dictated by the neutrino-neutrino interaction strength. Also, unlike other collective neutrino oscillation phenomena, fast flavor conversions can develop in the limit of vanishing vacuum frequency~\cite{Chakraborty:2016lct}.
Fast flavor conversion is triggered by the presence of crossings in the  angular distribution of the electron lepton number (ELN)~\cite{Izaguirre:2016gsx,Morinaga:2021vmc,Dasgupta:2021gfs}. The fact that ELN crossings have been found to occur in the decoupling region, see e.g.~Refs.~\cite{Shalgar:2019kzy, Morinaga:2019wsv, Nagakura:2019sig, Glas:2019ijo, Abbar:2019zoq, Abbar:2020qpi, Nagakura:2021hyb} and that their occurrence is favored  when the number densities of the electron flavors are comparable~\cite{Shalgar:2019kzy}, hints that collisions may have a crucial impact~\cite{Shalgar:2019kzy,Brandt:2010xa}, while also affecting the development of flavor conversion~\cite{Johns:2021qby,Sigl:2021tmj,Capozzi:2018clo,Shalgar:2020wcx,Sasaki:2021zld,Martin:2021xyl}.

In the region where the decoupling of neutrinos begins and the self-interaction potential is of $\mathcal{O}(10^5)$~km$^{-1}$, the typical frequency associated with fast flavor conversions is a few orders of magnitude smaller, i.e., $\mathcal{O}(10^2$--$10^3)$~km$^{-1}$. In the supernova neutrino decoupling region, Ref.~\cite{Shalgar:2020wcx} pointed out that collisions strongly affect the flavor evolution, enhancing flavor conversion; while the classically predicted collisional damping of flavor conversion only occurs for much larger collisional strengths. This counter-intuitive result is in agreement with the findings of  Ref.~\cite{Sasaki:2021zld}. In contrast, Ref.~\cite{Martin:2021xyl}, considering a substantially different system, pointed out that fast conversion is damped by collisions. 

This work aims to shed light on the impact of collisions on  flavor evolution found in Refs.~\cite{Shalgar:2020wcx,Martin:2021xyl}. In order to do so, similar to Ref.~\cite{Sasaki:2021zld}, we rely on  polarization vector formalism. However, we expand on it as we consider two discrete ELN distributions in angle; one approximately uniform and the other one strongly forward peaked. We explore the interplay between flavor conversion and collisions occurring in these two scenarios. By relying on the normal mode analysis~\cite{Banerjee:2011fj,Padilla-Gay:2021haz}, we analytically compute the growth rate and initial precession frequency and demonstrate how the enhancement of flavor conversions, in the presence of collisions, can be explained as an ``adiabatic'' enhancement.

This paper is organized as follows:
In \sref{sec:threebin} we introduce the three bin model and outline our assumptions. Subsequently, fast flavor conversions without collisions are analyzed in \sref{sec:nocollisions} for our three bin model, where we present the full solution and a linear stability analysis. Collisions are introduced in \sref{sec:collisions}, where the full solution is presented and interpreted using our knowledge on the initial linear regime. Finally, discussion and conclusions are found in \sref{sec:discussion} and \ref{sec:conclusions}.

%%%%%%%%%%%%%%%%%%%%%%%%%%%%%%%%%%%%%%%%%%%%%%%%%%%%%%%%%%%%%%%%%%%%%%%%%%%%%%%%%%%%%%%%%%%%%%%%
\section{Three bin neutrino model}
\label{sec:threebin}
In order to understand the conditions under which collisions lead to enhancement or suppression of fast flavor conversions, we consider a simple model consisting of neutrinos distributed along three angle bins. This allows us to easily track the evolution of each angle bin.
In the following, we describe the system used in this work, introduce its equations of motion, and set up the notation.

\subsection{Equations of motion}
For the sake of simplicity, we use the two flavor approximation along with the assumption of a single neutrino energy.
In the two flavor basis, ($\nu_e$, $\nu_x$), where $\nu_x$ represents a linear combination of $\nu_\mu$ and $\nu_\tau$, the flavor content of an homogeneous neutrino ensemble can be expressed in terms of a $2 \times 2$ density matrices
\begin{equation}
  \label{eq:rho}
  \rho =
  \begin{pmatrix}
    \rho_{ee} & \rho_{ex} \\ \rho_{xe} & \rho_{xx}
  \end{pmatrix}
  \;, \qquad
  \bar{\rho} =
  \begin{pmatrix}
    \bar{\rho}_{ee} & \bar{\rho}_{ex} \\ \bar{\rho}_{xe} & \bar{\rho}_{xx}
  \end{pmatrix} \;
\end{equation}
at each point in phase space. 
It should be noted that in the context of fast flavor conversion, the two flavor approximation is not justified in the non-linear regime; however, the two flavor approximation is insightful due to its simplicity~\cite{Shalgar:2021wlj,Capozzi:2020kge,Chakraborty:2019wxe}.
The evolution of this system  is described by the quantum kinetic equations~\cite{Sigl:1992fn, Vlasenko:2013fja, Volpe:2013uxl}:
\begin{equation}
  \label{eq:QKE}
  \dot\rho(\vec{x},\vec{p}) = - i [ H_{\nu\nu}(\vec{x},\vec{p}), \rho(\vec{x},\vec{p}) ] + \mathcal{C}(\rho,\bar{\rho})\ ,
\end{equation}
where   $\vec{x}$ indicates the location of the neutrino field and $\vec{p}$ its momentum. 
In dense environments, the Hamiltonian for a neutrino with momentum $\vec{p}$ is $H_{\nu\nu}(\vec{p}) = \mu \int d\vec{p^{\prime}} [\rho(\vec{x},\vec{p^\prime})- \bar\rho(\vec{x},\vec{p^\prime})] (1- \vec{v} \cdot \vec{v^{\prime}})$, where $\mu$ is the self-interaction potential, which is proportional to the neutrino number density. 
The velocity vectors for the neutrino under consideration and the neutrinos in the medium are represented by $\vec{v}=\vec{p}/|\vec{p}|$ and $\vec{v^{\prime}}=\vec{p^{\prime}}/|\vec{p^{\prime}}|$, respectively.  Since we intend to focus on fast flavor conversion, we neglect the vacuum term in the Hamiltonian for simplicity, albeit fast flavor conversion can be affected by a non-vanishing vacuum frequency~\cite{Shalgar:2020xns}.

For the sake of simplicity, we assume homogeneity of the neutrino gas, azimuthal symmetry (hence the neutrino field is characterized by the polar angle, $\theta$), as well as  mono-energetic neutrinos. However, see Refs.~\cite{Shalgar:2021wlj, Shalgar:2021oko, Shalgar:2021wlj, Shalgar:2020xns, Shalgar:2019qwg, Martin:2019kgi, Richers:2021xtf} for dedicated work exploring the impact of the simplifying assumptions adopted here.

The local isomorphism between the SU(2) and SO(3) groups allows us to represent the $2\times 2$ density matrices as  three-dimensional Bloch vectors called  ``polarization vectors''~\footnote{Throughout this paper, we  use arrows to denote vectors in the real space and bold fonts to denote vectors in the flavor space.}: 
\begin{equation}
  \label{eq:polvec}
  \rho = \frac{1}{2}(P_0 +\mathbf{P} \cdot \bm{\sigma})\ ,
\end{equation}
where $\bm{\sigma} = (\sigma_x, \sigma_y, \sigma_z)^T$ is the vector of Pauli matrices. Similarly, the density matrix for antineutrinos is expressed as a function of the polarization vector $\bar{\mathbf{P}}$. 
The Hamiltonian can be written in terms of the ``potential vector'' by means of  the following relation:
\begin{equation}
  \label{eq:potvec}
  H_{\nu\nu} = \frac{1}{2}(V_{\nu\nu,0} +\mathbf{V}_{\nu\nu} \cdot \bm{\sigma})\;.
\end{equation}
The equations of motion for flavor evolution introduced in \eref{eq:QKE} can be re-written by using \eref{eq:polvec} and \eref{eq:potvec}:
\begin{equation}
  \label{eq:eom3}
  \begin{aligned}
    \dot{\mathbf{P}}(u_{i}) &= \mathbf{V}_{\nu\nu}(u_i) \times \mathbf{P}(u_{i}) \;,\\
    \dot{\bar{\mathbf{P}}}(u_{i}) &= \mathbf{V}_{\nu\nu}(u_i) \times \bar{\mathbf{P}}(u_{i}) \;,
  \end{aligned}
\end{equation}
where $u_i = \cos \theta_i$. It should be noted that the angle bins $u_i$ are not equally spaced and do not have the same bin-widths. We use $\Delta u_i$ to denote the bin-width of the $i$'th bin.
The part of $\mathbf{V}_{\nu\nu}$ that is independent of $u_i$ can be removed by going to a rotating frame (RF)~\cite{Duan:2005cp, Duan:2006an}. After summing over $j$, one obtains 
\begin{align}
  \label{eq:Vnunu}
  \mathbf{V}_{\nu\nu}^{\rm RF}(u_i) = -\mu u_i \mathbf{D}_1^{\rm RF} \;,
\end{align}
where $\mathbf{D}_1^{\rm RF}$ is the first moment of the difference between neutrino and antineutrino polarization vectors. In general:
\begin{equation}
  \label{eq:Dndef}
  \mathbf{D}_n = \sum_{j} [\mathbf{P}(u_j) - \bar{\mathbf{P}}(u_j)] u_j^n \Delta u_j \;.
\end{equation}

It is also useful to express the equations of motion through the sums and differences of the polarization vectors:
\begin{equation}
  \label{eq:Ppm}
  \mathbf{S}(u_i) = \mathbf{P}(u_i) + \bar{\mathbf{P}}(u_i) \;, \qquad   \mathbf{D}(u_i) = \mathbf{P}(u_i) - \bar{\mathbf{P}}(u_i) \;.
\end{equation}
Hence, in the rotating frame, \eref{eq:eom3} gives:
\begin{equation}
  \label{eq:eompm}
  \begin{aligned}
     \dot{\mathbf{S}}^{\rm RF}(u_{i}) &= \mu u_i \mathbf{S}^{\rm RF}(u_{i}) \times \mathbf{D}^{\rm RF}_1\;,\\
     \dot{\mathbf{D}}^{\rm RF}(u_{i}) &= \mu u_i \mathbf{D}^{\rm RF}(u_{i}) \times \mathbf{D}^{\rm RF}_1\;,
  \end{aligned}
\end{equation}
highlighting  that $\mathbf{D}^{\rm RF}(u_i)$ evolves independently of $\mathbf{S}^{\rm RF}(u_i)$~\cite{Padilla-Gay:2021haz}. 

\subsection{System setup}
\label{subsec:setup}
\begin{figure*}
  \centering
  \includegraphics[width=\columnwidth]{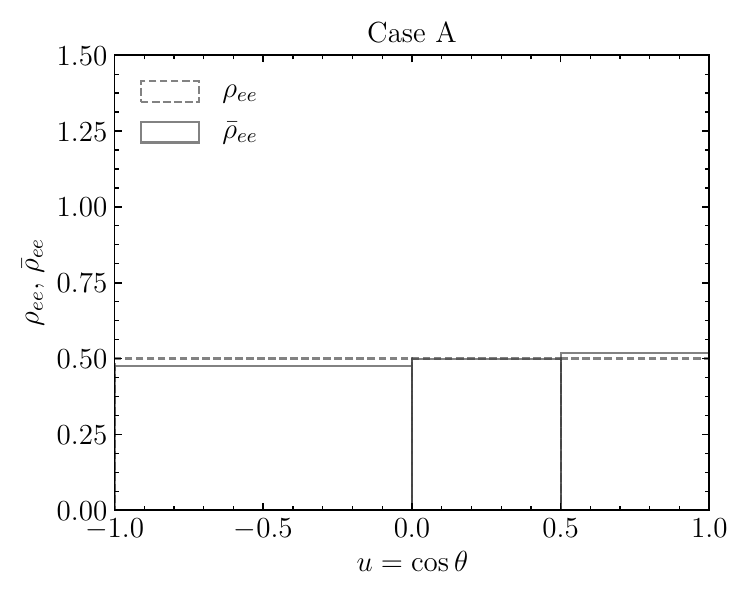}
  \includegraphics[width=\columnwidth]{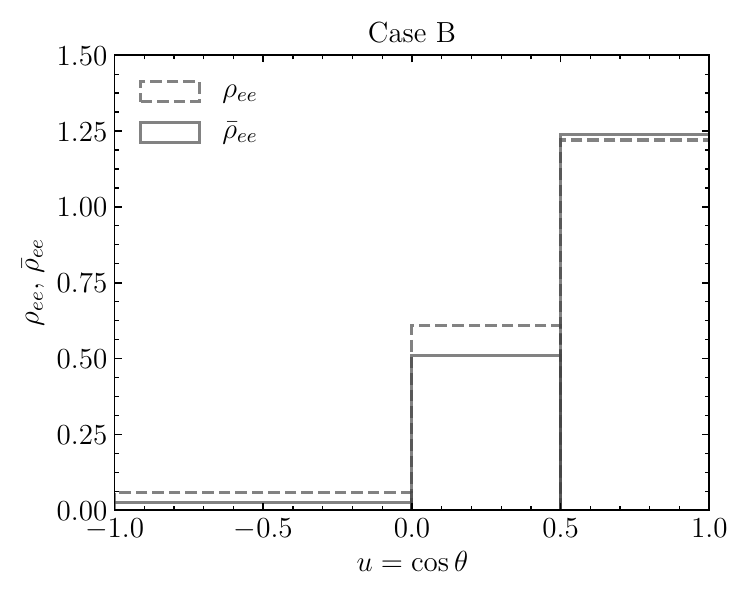}
  \caption{Three mode angular distributions of $\nu_e$ (dashed lines) and $\bar\nu_e$ (solid lines)  for case A and case B (see also \tref{tab:udist}).}
  \label{fig:udist}
\end{figure*}

\begin{table}
  \caption{Initial angular distributions for the electron flavors for cases A and B. $\rho_{xx}(u_i)=\bar{\rho}_{xx}(u_i)=0$ initially for all $u_i$ in both cases.}
  \centering
  \begin{tabular}{l c c c}\hline\hline
    Bin & $u_1$ & $u_2$ & $u_3$\\\hline\\[-0.2cm]
    $u_i$ & $-0.50$ & $0.25$ & $0.75$\\
    $\Delta u_i$ & $1.00$ & $0.50$ & $0.50$ \\\hline\\[-0.2cm]
    Case A & & &\\
    $\rho_{ee}(u_i)$ & $0.500$ & $0.500$ & $0.500$\\
    $\bar{\rho}_{ee}(u_i)$ & $0.475$ & $0.498$ & $0.517$\\\hline\\[-0.2cm]
    Case B & & &\\
    $\rho_{ee}(u_i)$ & $0.058$ & $0.609$ & $1.220$\\
    $\bar{\rho}_{ee}(u_i)$ & $0.025$ & $0.509$ & $1.239$\\\hline\hline
  \end{tabular}
  \label{tab:udist}
\end{table}

For an azimuthally symmetric system, the angular distribution can be described in terms of $u=\cos\theta$. We represent the angular distribution by means of three bins: one bin in the backward direction ($u_1$) and two bins in the forward direction ($u_2$ and $u_3$). 
We consider two cases: case A represents a relatively uniform angular distribution; it resembles the one previously adopted in Ref.~\cite{Shalgar:2020wcx} and  is shown  in the left panel of \fref{fig:udist}. The second benchmark distribution  (``case B'') is more forward peaked than case A and mimics the angular distribution adopted in Ref.~\cite{Martin:2021xyl}. It is shown in the right panel of \fref{fig:udist}. The angular distributions are reported in \tref{tab:udist}, where $u_i$ and $\Delta u_i$ are also listed. 

The three bins in \fref{fig:udist} exhibit a crossing in the forward direction for both case A and B. Furthermore, it has been demonstrated that in the absence of collisions, three angular bins are sufficient to characterize the evolution of fast flavor conversions in the homogeneous case~\cite{Padilla-Gay:2021haz}. As we will later demonstrate, also with collisions, three bins qualitatively reproduce the results from continuous distributions for the benchmark cases we consider.

\section{Fast flavor conversion in the absence of collisions}
\label{sec:nocollisions}
In this section, we  compare the flavor evolution in the non-linear regime for cases A and B in the absence of collisions. We then introduce the linearized equations of motion and compare the full solutions to the eigenvalues and eigenvectors computed through the linear normal mode analysis.
\subsection{Non linear flavor evolution}
\label{subsec:nonlinear}
We  assume  $\mu = 10^5\;{\rm km}^{-1}$. 
In order to trigger flavor conversion, we take the first component of the polarization vector to be $P_x(u_i, t=0) = 10^{-8}$. 
In order to quantify the amount of flavor conversion, we compute the conversion probability 
\begin{equation}
  \label{eq:Pex}
  P_{ex} = \frac{1}{2}\left(1 - \frac{ \sum_i \Delta u_i  P_z(u_i)}{\sum_i \Delta u_i P_{0}(u_i)}\right) \;,
\end{equation}
which is unchanged by going to the rotating frame.
In the absence of collisions, $P_0$ is constant and determined from the initial conditions.

\begin{figure*}
  \centering
  \includegraphics[width=\columnwidth]{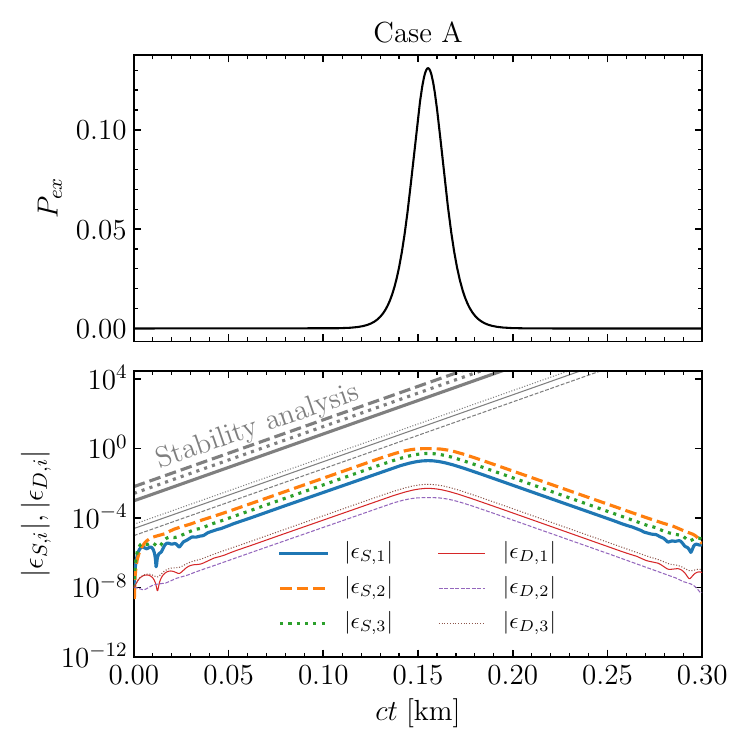}
  \includegraphics[width=\columnwidth]{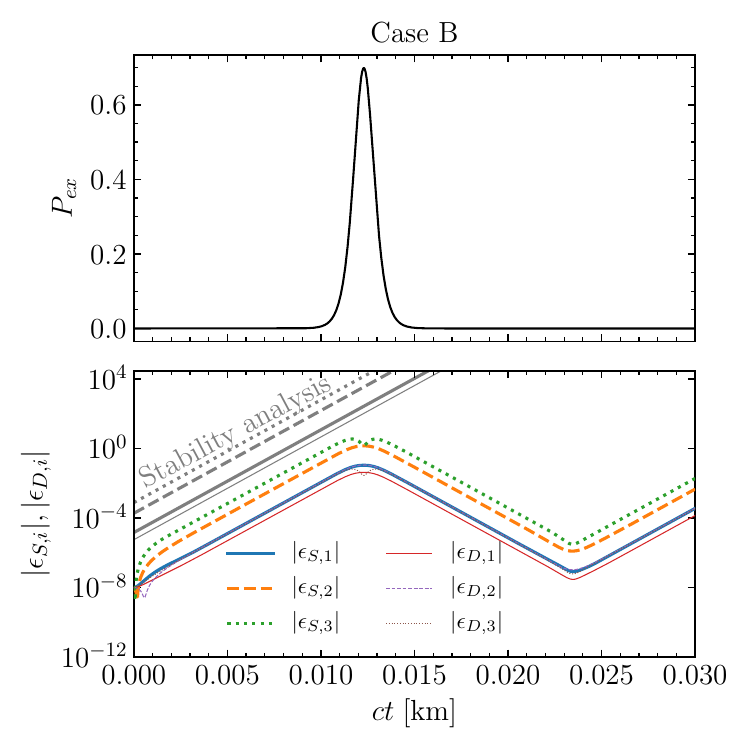}
  \caption{Fast flavor conversion for the three bin neutrino model for cases A (left panels) and B (right panels) in the absence of collisions. {\it Top panels:} Conversion probability (see \eref{eq:Pex}) as a function of the distance. Case A shows overall less flavor conversion than case B, moreover the timescale over which flavor conversion occurs is different in the two cases. {\it Bottom panels:} Components of 
  $\epsilon_{S,i}$  and $\epsilon_{D,i}$ (see \eref{eq:eps}) as functions of the distance and for each of the three angular bins compared to the growth rate and eigenvector from linear stability analysis (Eqs.~(\ref{eq:eval}) and (\ref{eq:evec})).  It should be noted that  an overall offset in the results of the linear stability analysis (gray lines) has been introduced for the sake of readability.  Note that the imaginary part of the eigenvalue for case B is larger than the one of case A. The full solution and the eigenvectors are in good agreement for cases A and B. In the lower right panel, the curves for $|\epsilon_{D,2}|$ and $|\epsilon_{D,3}|$ are both on top of $|\epsilon_{S,1}|$.}
  \label{fig:onset}
\end{figure*}

The top panels of \fref{fig:onset} show the conversion probability $P_{ex}$ as a function of the distance (time) for cases A (on the left) and B (on the right). 
 As for case A,  the maximal conversion probability is larger in the three bin model than in the case with continuous distributions (see Fig.~2 of Ref.~\cite{Shalgar:2020wcx}), and the conversions start earlier, but the qualitative behavior is similar despite the significant simplification of using three bins only.
 As for case B, the top right panel of Fig.~\ref{fig:onset}  shows that the time scale of flavor conversion is approximately ten times faster  than for case A (see the right panels of Fig.~3 in~\cite{Martin:2019gxb} for comparison). In addition,  the maximal flavor conversion in case B is much larger than in case A. Apart from this, the behavior of cases A and B  is qualitatively similar.

%%%%%%%%%%%%%%%%%%%%%%%%%%%%%%%%%%%%%%%%%
\subsection{Insights from the linearized equations of motion}
The initial flavor evolution during the linear phase can be explored through  a linearized set of equations~\cite{Banerjee:2011fj}. In this context, the stability of the system can be addressed by calculating the growth rate, and the associated normal modes give valuable information regarding the characteristics of the  flavor instability~\cite{Izaguirre:2016gsx,Padilla-Gay:2021haz}.

The initial flavor state consists of almost pure $\nu_e$ and $\bar{\nu}_e$. Hence the off-diagonal parts of the density matrices are small, and we can follow the initial evolution by expanding the equations in terms of these off-diagonal parts. As demonstrated in \eref{eq:eompm}, $\mathbf{D}^{\rm RF}(u_i)$ decouples from $\mathbf{S}^{\rm RF}(u_i)$, and for this reason we use $\mathbf{D}^{\rm RF}(u_i)$ and $\mathbf{S}^{\rm RF}(u_i)$ as a basis for the stability analysis.
In order to follow the initial onset of flavor conversion, we introduce the following linear combinations:
\begin{equation}
  \label{eq:eps}
  \epsilon_{S,i} = S_x^{\rm RF}(u_i) - i S_y^{\rm RF}(u_i) \;, \quad \epsilon_{D,i} = D_x^{\rm RF}(u_i) - i D_y^{\rm RF}(u_i) \;.
\end{equation}
When linearizing \eref{eq:eompm}, we only keep the terms linear in $\epsilon_{S,i}$ and $\epsilon_{D,i}$.
Writing the small perturbation  in terms of two vectors, $\epsilon_S = (\epsilon_{S,1}, \epsilon_{S,2}, \epsilon_{S,3})^T$ and $\epsilon_D = (\epsilon_{D,1}, \epsilon_{D,2}, \epsilon_{D,3})^T$, we can express the linearization as a matrix equation:
\begin{equation}
  \label{eq:linMatrix}
  \begin{pmatrix}
    \dot{\epsilon}_S \\ \dot{\epsilon}_D
  \end{pmatrix} = -i M
  \begin{pmatrix}
    \epsilon_S \\ \epsilon_D
  \end{pmatrix} = -i 
  \begin{pmatrix}
    M_{SS} & M_{SD}\\ M_{DS} & M_{DD}
  \end{pmatrix}
  \begin{pmatrix}
    \epsilon_S \\ \epsilon_D
  \end{pmatrix} \;.
\end{equation}
Here $M_{DS}$ is zero, and the other three matrices are given by
\begin{align}
  \label{eq:MSS}
  M_{SS,ij} &= -\mu D_{1,z}^{\rm RF} u_i \delta_{ij} \;, \\
  \label{eq:MSD}
  M_{SD,ij} &=  \mu u_i S_z^{\rm RF}(u_i) u_j \Delta u_j \;,\\
  \label{eq:MDD}
  M_{DD,ij} &= -\mu D_{1,z}^{\rm RF} u_i \delta_{ij} + \mu u_i D_z^{\rm RF}(u_i) u_j \Delta u_j \;,
\end{align}
with $\delta_{ij}$ being the Kronecker delta. 

The solution of these equations is known to be collective in nature, hence we apply the following ansatz: 
\begin{equation}
  \label{eq:epsansatz}
  \epsilon_S(t) = \epsilon_S(0) \exp(-i\Omega t)\;, \quad
  \epsilon_D(t) = \epsilon_D(0) \exp(-i\Omega t)\;.
\end{equation}
Through the ansatz above, the time derivative in \eref{eq:linMatrix} can be evaluated, and the resulting equation can be recast as a homogeneous equation. 
This equation has a solution if and only if $\det(M - \Omega) = 0$. 
The imaginary part of the eigenvalue $\Omega$ will lead to exponentially growing or decaying solutions, while the real part corresponds to oscillations of $\epsilon_{S/D}$ in the complex plane. 
Hence, a solution for $\Omega$ that has a positive imaginary component implies a flavor instability, which can lead to significant flavor transformation.

The eigenvalues can be determined analytically since $M_{DS}=0$. The eigenvalues from $M_{SS}$ are all real, and since exponentially growing solutions have complex eigenvalues, we focus on $M_{DD}$. Here we find the solution:
\begin{equation}
  \label{eq:eval}
  \begin{aligned}
    \Omega  &= -\frac{\mu}{2} \left(\sum_{i} u_i  D_{1,z}^{\rm RF} - D_{2,z}^{\rm RF} \right)\\
    &\quad\pm i \;\frac{\mu}{2} \sqrt{ 4 u_1 u_2 u_3 D_{0,z}^{\rm RF} D_{1,z}^{\rm RF} - \left(\sum_iu_i D_{1,z}^{\rm RF} - D_{2,z}^{\rm RF}\right)^2} \;,
\end{aligned}
\end{equation}
which is a complex conjugate pair if the argument in the square root is positive. The moments $D_{n,z}^{\rm RF}$ are defined according to \eref{eq:Dndef}. 
The exponentially growing solution dominates over all other solutions within a short time span, and hence the eigenvalue with the largest imaginary part determines the degree of instability. 

The following  eigenvalues are obtained for cases A and  B (see \tref{tab:udist}):
\begin{eqnarray}
  \label{eq:eval_numA}
  \Omega_A &\approx& (536 \pm 88\; i)~{\rm km}^{-1}\;,\\
  \label{eq:eval_numB}
  \Omega_B &\approx& (653 \pm 1361 \; i)~{\rm km}^{-1}\;.
\end{eqnarray}
 The imaginary part of the eigenvalue for case B is more than ten times larger  than for case A, as also reflected in the different time scales in the left and right panels of \fref{fig:onset}.

The vectors $\epsilon_S(0)$ and $\epsilon_D(0)$, the normal modes, are given by the eigenvector corresponding to the eigenvalue and can be determined from Eqs.~(\ref{eq:linMatrix})-(\ref{eq:MDD}):
\begin{equation}
  \label{eq:evec}
  \begin{pmatrix}
    \epsilon_{S}(0)\\ \epsilon_{D}(0)
  \end{pmatrix}
  \propto \frac{u_i }{\Omega + \mu u_i D_{1,z}^{\rm RF}}
  \begin{pmatrix}
    S_z^{\rm RF}(u_i)\\ D_z^{\rm RF}(u_i)
  \end{pmatrix}
  \;,
\end{equation}
where $S_z^{\rm RF}(u_i)$ and $D_z^{\rm RF}(u_i)$ are vectors in $i$.
The initial conditions given in \sref{subsec:setup} and \ref{subsec:nonlinear} determine how each normal mode is populated.

The results from the linearized equations are compared to the full solution in the lower panels of \fref{fig:onset} for cases A and B. For case A and  $ct<0.02$~km (left panel), a few oscillations are visible before the exponentially growing solution starts to dominate. For $0.02 < ct < 0.14$, there is an excellent agreement between the full solution (colored lines) and the growth rate determined from the stability analysis (gray lines, offset for readability) since the slopes of the lines agree very well.
In addition, the ratios between the different lines in the full solution are reproduced nicely by the eigenvectors from the stability analysis, thus confirming that the expected normal modes are dominating. 

The results for case B (right lower panel of \fref{fig:onset}) show a similarly good agreement between the full solution and the eigenvectors. In this case, the exponentially growing solution dominates for $0.002{\;\rm km} < ct < 0.011$ km.

The eigenvector also captures well the angular dependence of the neutrino flavor conversions in the non-linear regime for both cases, as it is evident from the comparison between the results of the linear stability analysis and the full solution. For case B, the conversion probability becomes so large that $|\epsilon_{S,i}|$ and $|\epsilon_{D,i}|$ show a small dip at $ct = 0.0125{\;\rm km}$ in the lower right panel of \fref{fig:onset}. 
It should be noted that in \fref{fig:onset}, the number of neutrinos is conserved for each angle bin independently. As we will see in the following, the conservation of neutrino number for each angle bin is broken when momentum changing collisions are included in the computation; however, the total number of neutrinos is still conserved in our simplified model of collisions.

%%%%%%%%%%%%%%%%%%%%%%%%%%%%%%%%%%%%%%%%%%%%%%%%%%%%%%%%%%%%%%%%%%%%%%%%%%%%%%%%%%%%%%%%%%%%%%%%
\section{Flavor conversion in the presence of collisions}
\label{sec:collisions}
In this section, we explore how flavor conversion is affected by collisions in cases A and B. First, we explore the overall non-linear flavor evolution, and then we focus on information carried by the linearized equations of motion.
In this section, we use the vectors $\mathbf{P}$ and $\bar{\mathbf{P}}$ to distinguish between neutrinos and antineutrinos.

\subsection{Non linear flavor evolution}
\begin{figure*}[tbp]
  \centering
  \includegraphics[width=\columnwidth]{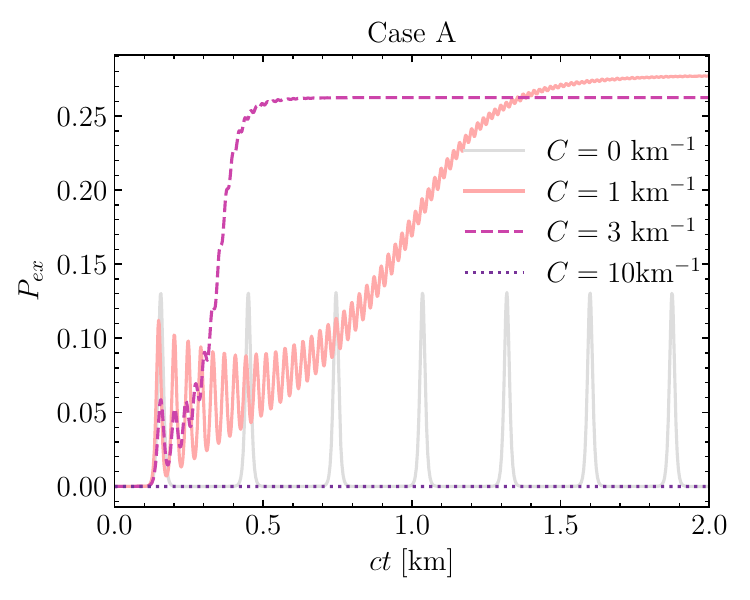}
  \includegraphics[width=\columnwidth]{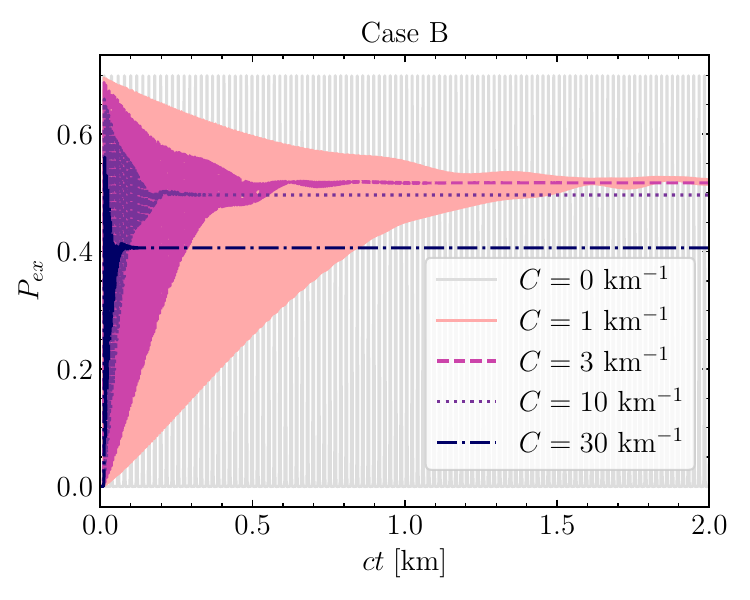}
  \caption{Fast flavor conversion for the three bin neutrino model for cases A (left panel) and B (right panel) in the presence of collisions. The conversion probability is shown for $C = 1$, $3$, and $10\;{\rm km}^{-1}$ for case A and $C = 1$, $3$, $10$ and $30\;{\rm km}^{-1}$ for case B. For both configurations, the  no-collision case ($C=0\;{\rm km}^{-1}$) is plotted in gray to guide the eye. For case A, intermediate values of $C$ lead to an enhancement of flavor conversion and larger values of  $C$ are responsible for suppressing flavor conversion. On the other hand, all configurations with $C \neq 0$ show a suppression of flavor conversion for case B.}
  \label{fig:col_sample}
\end{figure*}

Recently, it has been pointed out that non-forward collisions may enhance as well as suppress fast pairwise mixing~\cite{Shalgar:2020wcx, Martin:2021xyl, Sasaki:2021zld}. 
Before we proceed, we would like to clarify what we mean by enhancement of flavor mixing. We consider the enhancement of the asymptotic conversion probability with respect to the maximal conversion probability encountered when collisions are absent. This convention has the advantage of being independent of the initial condition (taking, for example, the average conversion probability over time would introduce a significant dependence on the plateaus between conversions and hence on the size of the initial perturbation). In addition, we wish to be conservative in claiming enhancement; hence,  using the maximal conversion allows to achieve this goal. The mechanism responsible for the suppression is explained by two effects. First, frequent incoherent oscillations destroy coherence; second, the incoherent collisions  tend to smear out the ELN crossings necessary for flavor conversion to occur~\cite{Morinaga:2021vmc}.  As we demonstrate in this section, the enhancement can also be understood by relying on similar arguments.

The full collision term, $\mathcal{C}$ in \eref{eq:QKE}, is computationally expensive to evaluate. In the spirit of simplicity, we  adopt the approximation used in Ref.~\cite{Shalgar:2020wcx}, which only focused on direction changing collisions. For neutrinos, the following  collision term is added to \eref{eq:eom3} 
\begin{equation}
  \label{eq:C}
  \frac{C}{2} \left[\sum_{j} \mathbf{P}(u_j) \Delta u_j - 2 \mathbf{P}(u_{i})\right]\;.
\end{equation}
A similar expression for the collision term holds for antineutrinos, and it is also valid in the rotating frame. We use the scripted $\mathcal{C}$ in \eref{eq:QKE} to denote the collision term and the unscripted $C$ to denote the parameter that encapsulates the strength of the collision term.

The left panel of \fref{fig:col_sample} shows the conversion probability for case A as a function of the distance. A clear enhancement of flavor conversion is visible for moderate values of $C$. Note that, comparing the solution with $C = 1$ km$^{-1}$ to the results of Fig.~2 of Ref.~\cite{Shalgar:2020wcx}, our results are in good  qualitative  agreement, but quantitatively they are not directly comparable because of the  three bin model adopted in this paper.
For larger values of $C$, the enhancement of flavor conversion occurs on shorter timescales;  for   $C \gtrsim 4.5$ km$^{-1}$,  flavor conversion is suppressed. 

On the contrary, there is no enhancement of flavor conversion for case B shown in the right panel of \fref{fig:col_sample}. For this configuration, even the lowest value of $C$ gives a suppression of the initial flavor conversion probability;  as $C$ is gradually increased to $30\;{\rm km}^{-1}$, the asymptotic value of the conversion probability decreases gradually. As $C$ increases, the conversion probability reaches a steady configuration at smaller distances.

\begin{figure*}
  \centering
  \includegraphics[width=\columnwidth]{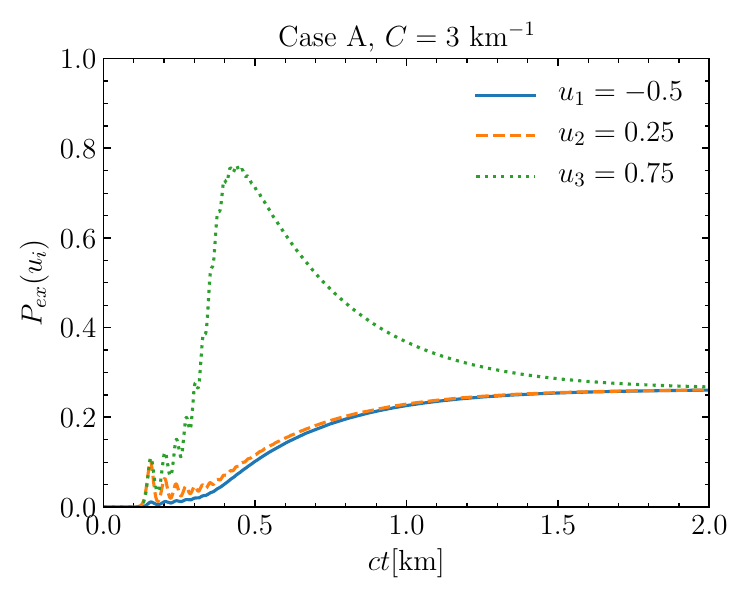}
  \includegraphics[width=\columnwidth]{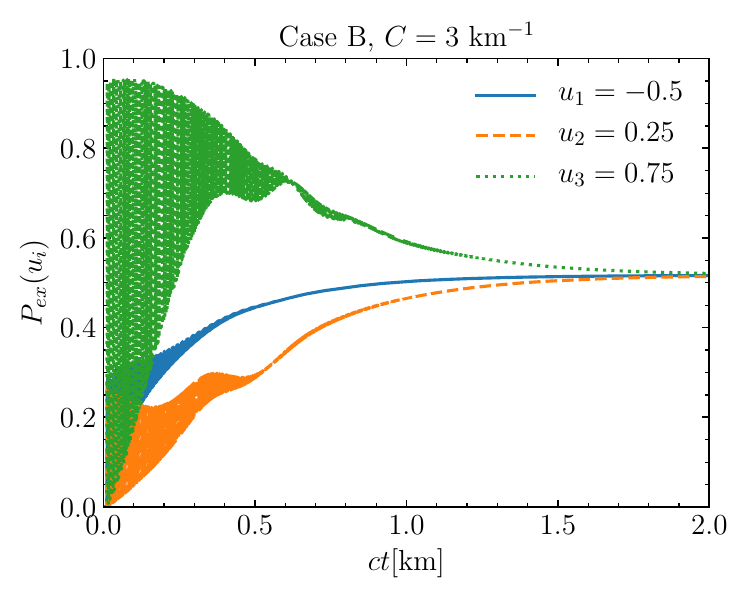}
  \caption{The transition probability (see Eq.~\ref{eq:Pexu}) due to fast flavor conversion and collisions for each of the three angular bins for cases A (left) and B (right) in the presence of collisions ($C=3 \; {\rm km}^{-1}$). In case A, flavor conversion sweeps from $u_2$ to $u_3$ and significantly develops in $u_3$ until $ct>0.5$ km. On the other hand, in case B, flavor conversion starts in $u_3$ and predominantly occurs in  this angular bin, only slightly affecting flavor in the bins centered on $u_1$ and $u_2$. Collisions dominate at $ct>0.7$ in both cases.}
  \label{fig:col_bins}
\end{figure*}

In order to investigate the origin of the opposite effect of collisions on the survival probability for cases A and B, we show in the \fref{fig:col_bins} the conversion probability in each of the three angular bins for cases A (left) and B (right):
\begin{equation}
  \label{eq:Pexu}
  P_{ex}(u_i) = \frac{1}{2}\left( 1 - \frac{ P_z(u_i)}{P_{0}(u_i)} \right)\;,
\end{equation}
where $P_0(u_i)$ as a function of time is given by
\begin{equation}
  \label{eq:P0u}
  P_0(u_i,t) = P_0(u_i,0) e^{-Ct} +  \left(1-e^{-Ct}\right)\frac{\sum_j P_0(u_j,0)\Delta u_j}{\sum_k \Delta u_k} \;. 
\end{equation}
A similar definition holds in the rotating frame.

The enhancement of the flavor conversion probability in case A is due to the fact that flavor conversion sweeps  from one part of the angular distribution to another, as shown in the left panel of  \fref{fig:col_bins}. Flavor conversion  initially takes place in the bin centered on $u_2=0.25$ and then gradually moves to the bin centered on $u_3=0.75$. The $u_3$ bin reaches its enhanced maximal value at $ct \sim 0.45$ km, where the total flavor conversion probability also reaches its maximal and asymptotic value (see also the left panel of \fref{fig:col_sample}). Finally, at $ct>0.5$ km, conversions cease and collisions start to dominate, equilibrating all angular bins. A similar trend is observed in Refs.~\cite{Shalgar:2020wcx, Sasaki:2021zld} for continuous angular distributions.

For case B, the right panel of \fref{fig:col_bins} shows that flavor conversion is the largest in the bin centered on $u_3=0.75$ throughout the simulation duration, and the conversion probability of $u_3$ is the largest at $ct \sim 0$ km before collisions have had an impact. Beyond $ct=1$ km, flavor oscillation occurs with  smaller amplitude, and collisions tend to drive all bins towards the same $P_{ex, S}$. Although it is clear that flavor conversion does not remain localized to one angular bin in case A, the mechanism responsible for this effect is not immediately obvious. One would expect that flavor conversion primarily occurs  close to the ELN crossing~\cite{Tamborra:2020cul} and, as the ELN crossing sweeps through the angular distribution because of collisions, flavor conversion could move along.

%%%%%%%%%%%%%%%%%%%%%%%%%%%%%%%%%%%%%%%%%
\subsection{Flavor evolution in the eigenframe}
\label{subsec:eigenframe}

\begin{figure*}[tbp]
  \centering
  \includegraphics[width=\textwidth]{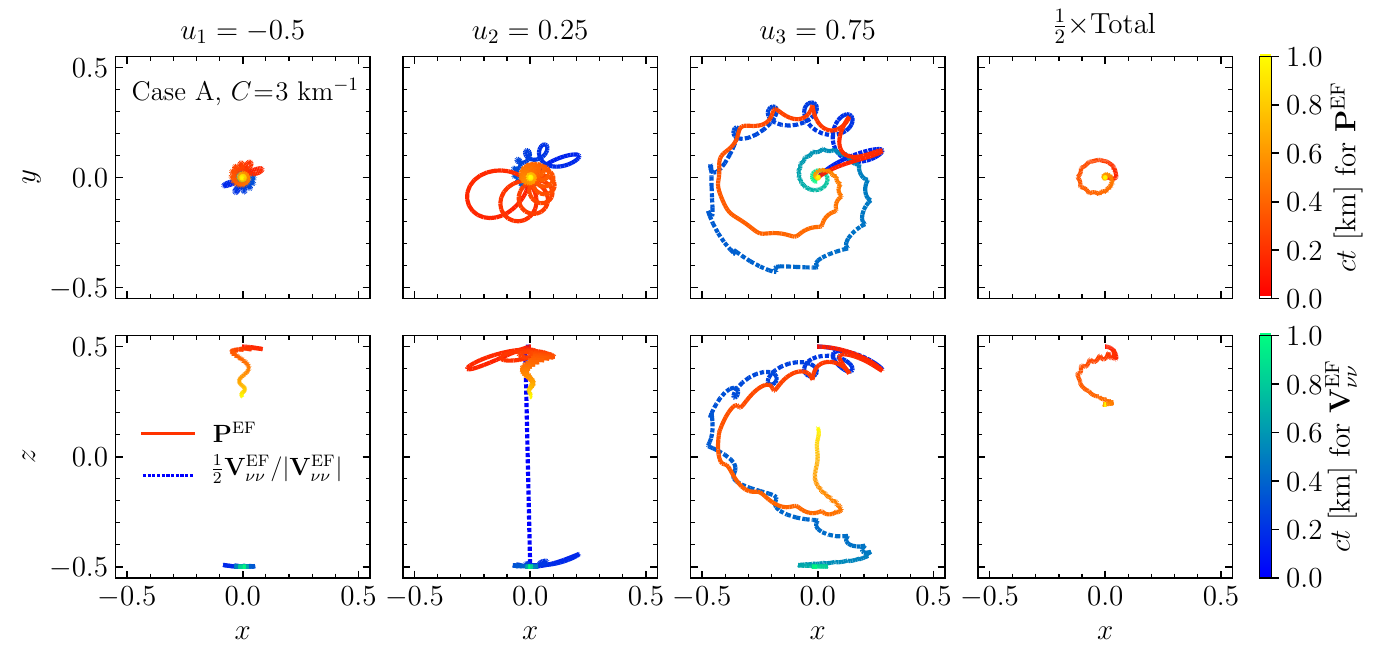}
  \caption{Evolution of the polarization vectors (solid lines with orange hues) and normalized potential vectors (dashed lines with blue hues) linked to the three angular bins (first three columns from left) and the   total polarization vector (panels on the right) for case A in the eigenframe and with $C = 3\; {\rm km}^{-1}$. The total polarization vector is calculated by integrating over the angular distribution and dividing by 2 for improved readability. The temporal evolution is represented by the color gradient, which becomes lighter as time increases. The top panels show a projection in the  $x$--$y$ plane (top view), while the bottom panels show a projection in the $x$--$z$ plane (side view).
    $\mathbf{P}^{\rm EF}$ and $\mathbf{V}_{\nu\nu}^{\rm EF}$ are parallel or anti-parallel for most of the evolution in all three bins. This ``adiabatic'' evolution explains the flavor conversion enhancement. 
An animated version of this figure is found in the \href{https://sid.erda.dk/share_redirect/eSC43aEyOq/index.html}{Supplemental Material}.}
  \label{fig:animation_still}
\end{figure*}

The eigenvalue that dominates the solution is a complex number whose imaginary part determines the growth rate in the linear regime, while the real part gives rise to oscillations in $P_x$ and $P_y$.
Such overall oscillations do not directly affect the flavor conversion seen in \fref{fig:col_bins}, and can be removed by going to a frame rotating with angular frequency ${\rm Re}(\Omega)$~\cite{Duan:2005cp, Duan:2006an}, which we will call the ``eigenframe'' (EF). 
The potential vector in the eigenframe is
\begin{equation}
  \label{eq:Vef}
  \mathbf{V}_{\nu\nu}^{\rm EF}(u_i) = -\mu u_i \mathbf{D}_1^{\rm EF} - (0,0,{\rm Re}(\Omega))^T\;,
\end{equation}
and the polarization vector in the eigenframe evolves according to
\begin{equation}
  \label{eq:eom4}
  \begin{aligned}
    \dot{\mathbf{P}}^{\rm EF}(u_{i}) &= \mathbf{V}_{\nu\nu}^{\rm EF}(u_i) \times \mathbf{P}^{\rm EF}(u_{i}) \\
    &\quad + \frac{C}{2} \left[\sum_{j} \mathbf{P}^{\rm EF}(u_j) \Delta u_j - 2 \mathbf{P}^{\rm EF}(u_{i})\right]\;,\\
    \dot{\bar{\mathbf{P}}}^{\rm EF}(u_{i}) &= \mathbf{V}_{\nu\nu}^{\rm EF}(u_i) \times \bar{\mathbf{P}}^{\rm EF}(u_{i}) \\
    &\quad + \frac{C}{2} \left[\sum_{j} \bar{\mathbf{P}}^{\rm EF}(u_j) \Delta u_j - 2 \bar{\mathbf{P}}^{\rm EF}(u_{i})\right]\;,\\
  \end{aligned}
\end{equation}
corresponding to Eqs. (\ref{eq:eom3}) and (\ref{eq:C}).

\begin{figure*}[tbp]
  \centering
  \includegraphics[width=\textwidth]{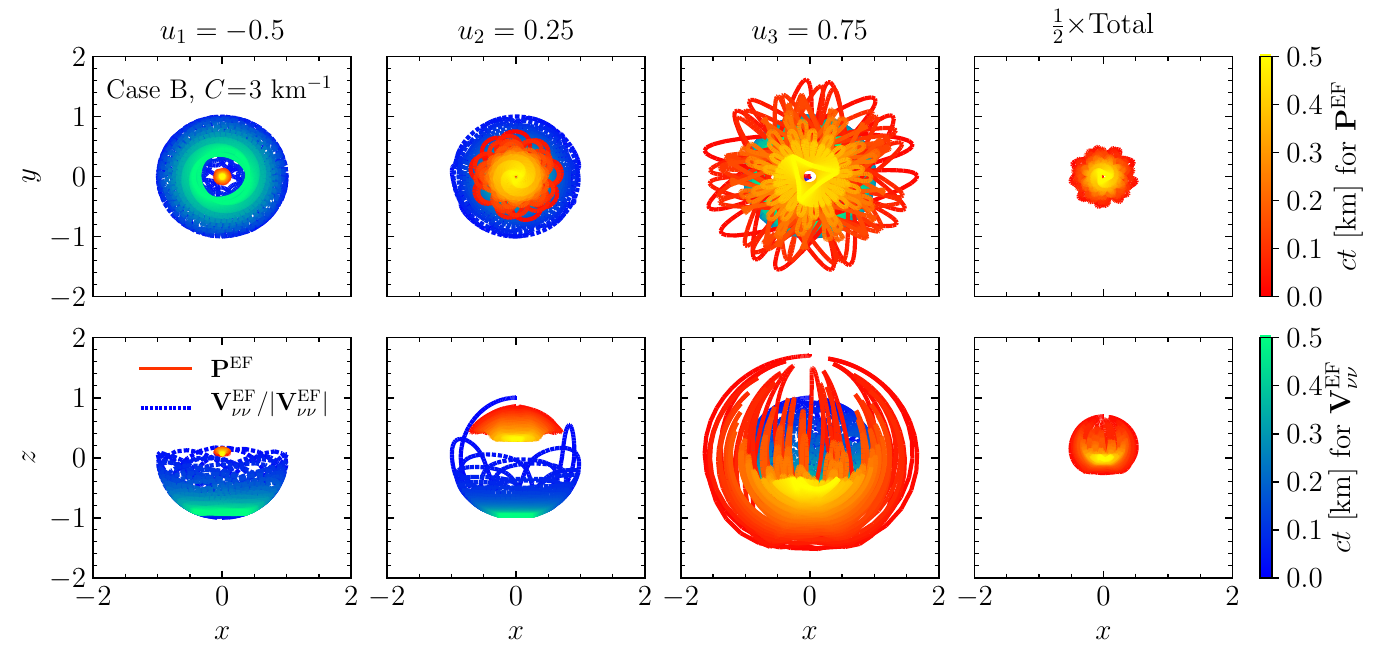}
  \caption{Same as \fref{fig:animation_still}, but for case B.  The evolution of $\mathbf{P}^{\rm EF}$ and $\mathbf{V}_{\nu\nu}^{\rm EF}$ is very different, and the evolution is not ``adiabatic'' as in \fref{fig:animation_still}. Also, there is no enhancement of flavor conversion. An animated version of this figure which focuses on the initial phase is found in the \href{https://sid.erda.dk/share_redirect/eSC43aEyOq/index.html}{Supplemental Material}.}
  \label{fig:animation_still_M}
\end{figure*}

The evolution of the flavor conversion for each angular bin in the eigenframe is displayed in \fref{fig:animation_still} for case A (see also the left panel of Fig.~\ref{fig:Vef} for the evolution of $V_{\nu\nu, z}^{\rm EF}$  and the \href{https://sid.erda.dk/share_redirect/eSC43aEyOq/index.html}{Supplemental Material}).
The polarization vector for all bins  rotates around the $z$-axis with a frequency of $\mathcal{O}(1\;{\rm km}^{-1})$. This is to be compared to ${\rm Re}(\Omega)/2\pi \sim \mathcal{O}(100\;{\rm km}^{-1})$. 
Hence, to a good approximation, the rotation frequency of the polarization vector continues to be dictated by the eigenvalue even in the non-linear regime (i.e., no rotation in the eigenframe).
The polarization vector associated with the first bin ($u_1=-0.5$) shows very little evolution as a function of time, precessing around the $z$-axis with small amplitude; similar to the potential
vector, although pointing in the opposite direction.  At late times, the polarization vector  relaxes towards the average over the three bins due to collisions.
The polarization vector corresponding to the second bin ($u_2=0.25$) initially precesses but also settles in the $z$-direction and relaxes towards the bin-averaged value as time passes. The corresponding potential vector starts out in the positive $z$-direction and very quickly swaps to negative values, where it stays without much change for the rest of the time of evolution.
The polarization vector linked to the third bin ($u_3=0.75$) starts out precessing and is gradually rotated from positive to negative $z$ along with the potential vector.
Towards the end of the considered temporal frame, the polarization vector detaches from the potential vector and relaxes towards the bin-averaged value like for the other bins.
Note that while $\mathbf{P}^{\rm EF}(u_i)$ and $\mathbf{V}_{\nu\nu}^{\rm EF}(u_i)$ are aligned for the third bin, they are anti-aligned for the first two bins. This means that $\mathbf{P}^{\rm EF}(u_i)$ and $\mathbf{V}_{\nu\nu}^{\rm EF}(u_i)$ for $i=1,2$ follow each other when the orange and blue tracks in the upper panels are rotated by $180^{\circ}$  with respect to each other.
The total polarization vector (right-most panels of  \fref{fig:animation_still}) shows a steady decrease from the initial value towards the final position, averaging out the differences between the three bins. The total potential (not shown) is not very good for characterizing the system and points towards the negative $z$ direction for the entire evolution. Hence it does not suggest that any flavor conversion should be present.

The evolution in the eigenframe for case B is shown in \fref{fig:animation_still_M} (see also the right panel of Fig.~\ref{fig:Vef} for the evolution of $V_z$ and the \href{https://sid.erda.dk/share_redirect/eSC43aEyOq/index.html}{Supplemental Material} for an animation of \fref{fig:animation_still_M}.). 
The polarization vector of the first bin ($u_1=-0.5$) starts out in the positive $z$ direction and stays in that hemisphere. $\mathbf{V}_{\nu\nu}^{\rm EF}(u_1)$ similarly starts out in the negative $z$ direction and remains in that hemisphere. Although the $u_1$ polarization vector and $\mathbf{V}_{\nu\nu}^{\rm EF}(u_1)$ start out being anti-parallel, they quickly break that configuration and evolve quite differently.
In the second bin, both $\mathbf{P}^{\rm EF}(u_2)$ and $\mathbf{V}_{\nu\nu}^{\rm EF}(u_2)$ start out in the positive $z$ direction. However, $\mathbf{V}_{\nu\nu}^{\rm EF}(u_2)$ rotates to negative $z$ in the very beginning and stays there. As for the first bin, the two vectors are neither aligned nor anti-aligned except for the initial configuration.
The third bin shows large amplitude oscillations, where both $\mathbf{P}^{\rm EF}(u_3)$ and $\mathbf{V}_{\nu\nu}^{\rm EF}(u_3)$ oscillate between positive and negative values of $z$. Again the two vectors show no alignment except for the initial state, and $\mathbf{V}_{\nu\nu}^{\rm EF}(u_3)$ rotates very quickly in the opposite direction of $\mathbf{P}^{\rm EF}(u_3)$.

Since the original angular distribution for case B is strongly forward peaked (see \fref{fig:udist}), the overall evolution of the polarization vector is dominated by $\mathbf{P}^{\rm EF}(u_3)$, while the other two polarization vectors corresponding to the remaining angular bins have much smaller initial values.
The effect of collisions is to damp flavor conversion and push all three polarization vectors towards isotropy, which in this case is very close to $\mathbf{P}^{\rm EF}=0$.

\subsection{Adiabaticity and alignment}
The fact that the $\mathbf{P}^{\rm EF}(u_3)$ very closely tracks the potential vector in case A suggests that the flavor conversion physics can be described as being adiabatic~\cite{Raffelt:2007xt}. 
On the contrary, case B shows no such tracking and the evolution appears to be non-adiabatic.
In order to gauge the degree of adiabaticity, we introduce the following parameter measuring the 
alignment between $\mathbf{P}^{\rm EF}(u_i)$ and $\mathbf{V}_{\nu\nu}^{\rm EF}(u_i)$: 
\begin{equation}
  \label{eq:alignment}
  a_i(t) = \frac{\mathbf{P}^{\rm EF}(u_i, t) \cdot \mathbf{V}_{\nu\nu}^{\rm EF}(u_i, t)}{|\mathbf{P}^{\rm EF}(u_i, t)| |\mathbf{V}_{\nu\nu}^{\rm EF}(u_i, t)|} \;,
\end{equation}
where the time dependence is indicated explicitly.
For $a_i(t)=1$, the vectors $\mathbf{P}^{\rm EF}(u_i,t)$ and $\mathbf{V}_{\nu\nu}^{\rm EF}(u_i,t)$ are aligned, while they are anti-aligned  for $a_i(t)=-1$.
The values of $|a_i|$ averaged over time are seen in \tref{tab:alignment}, where we  indeed find values very close to $1$ for case A and smaller than $1$ for case B. 
\begin{table}
  \centering
    \caption{Alignment parameters $|a_i|$ averaged from $ct=0\;{\rm km}$ through $1\;{\rm km}$ for $C=3\;{\rm km}^{-1}$ and $C=0\;{\rm km}^{-1}$. Values close to 1 indicate a high degree of adiabaticity, and $\left<|a_i|\right> > 0.99$ are indicated in bold.}
  \begin{tabular}{l c c c}\hline\hline
      & $\left<|a_1|\right>$ & $\left<|a_2|\right>$ & $\left<|a_3|\right>$\\\hline\\[-0.2cm]
    $C=3\;{\rm km}^{-1}$ & & &\\
    Case A & {\bf 1.000} & {\bf 0.998} & {\bf 0.994}\\
    Case B & 0.895 & 0.760 & 0.738\\\hline\\[-0.2cm]
    $C=0\;{\rm km}^{-1}$ & & &\\
    Case A & {\bf 1.000} & 0.891 & {\bf 1.000}\\
    Case B & 0.976 & 0.892 & 0.942\\\hline
    \hline
  \end{tabular}
  \label{tab:alignment}
\end{table}

The importance of the $C\rightarrow 0$ limit can be understood by looking at the asymptotic conversion probability in the left panel of \fref{fig:col_sample}. They show very little dependence on $C$ until it reaches the threshold where conversions are entirely suppressed. This implies that even a very small value of $C$ would show the same enhancement as $C=3\;{\rm km}^{-1}$ given large enough $t$. Similarly, the adiabaticity of the solution can be addressed in the $C\rightarrow 0$ limit.

Comparing the values for $C=3\;{\rm km}^{-1}$ and $C=0\;{\rm km}^{-1}$ in \tref{tab:alignment}, the pattern is similar for most bins as expected. However, there is one notable exception: The second bin ($\left<|a_2|\right>$) for case A shows significantly more alignment for $C=3\;{\rm km}^{-1}$ than for $C=0\;{\rm km}^{-1}$. The reason is that $V_{\nu\nu,z}^{\rm EF}(u_2)$ flips from positive to negative values at $t\sim 0$ for $C=3\;{\rm km}^{-1}$, thus changing the dynamics for the polarization vector of that angular bin. Before addressing this, we will look at an approximation for $V_{\nu\nu,z}^{\rm EF}$.

\begin{figure*}[tbp]
  \centering
  \includegraphics[width=\columnwidth]{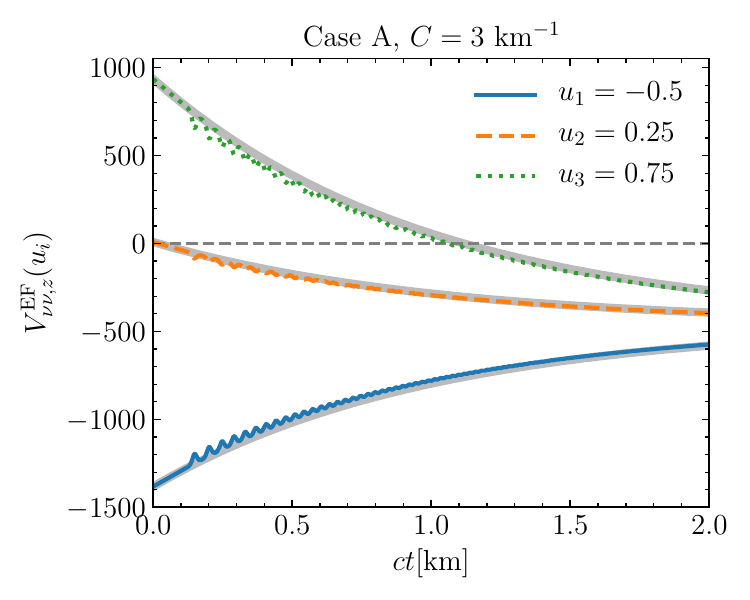}
  \includegraphics[width=\columnwidth]{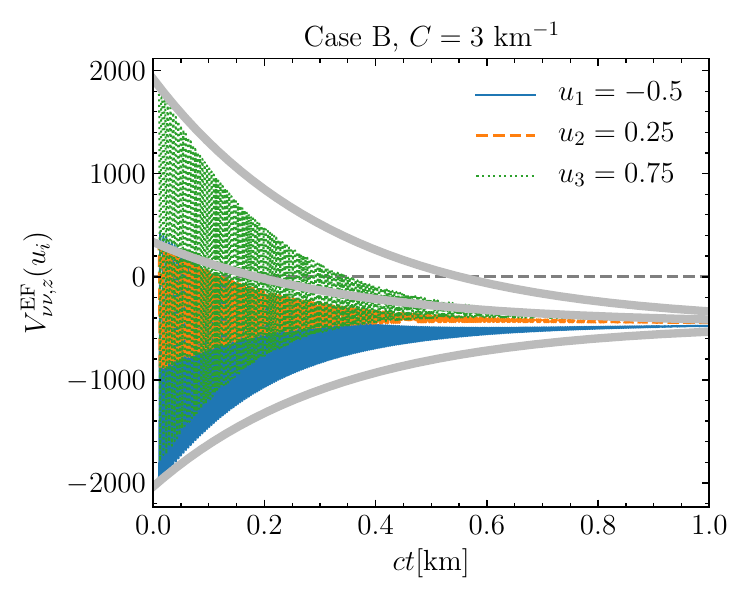}
  \caption{Evolution of the $z$ component of the potential in the eigenframe for the three angular bins in case A (left) and case B (right). The colored thin lines give the full solution, while the thick gray lines represent simple relaxation which is the evolution in the absence of flavor conversion (see \eref{eq:Vzrelax}). The full solution for $V_{\nu\nu, z}^{\rm EF}$ in case A is well approximated by the simple relaxation, while the full solution in case B is very different.}
  \label{fig:Vef}
\end{figure*}

In case A, the evolution of $V_{\nu\nu,z}^{\rm EF}$ can be approximated very well by neglecting the oscillation term in \eref{eq:eom4}. The result is an exponentially decaying solution:
\begin{align}
  V_{\nu\nu, z}^{\rm EF}(u_i,t) &= V_{\nu\nu, z}^{\rm EF}(u_i,0) e^{-Ct}\nonumber\\
  &\quad+  \left(1-e^{-Ct}\right)\frac{\sum_j V_{\nu\nu, z}^{\rm EF}(u_j,0)\Delta u_j}{\sum_k \Delta u_k} \;.   \label{eq:Vzrelax}
\end{align}
It is indicated by the thick gray lines in \fref{fig:Vef}. The full solutions are shown with colored lines.

Based on the alignment values in \tref{tab:alignment} for $C=0\;{\rm km}^{-1}$ and the inferred adiabaticity, we can make predictions for the behavior for the different bins in case A. 
  By combining adiabaticity with our approximate $V_{\nu\nu,z}^{\rm EF}(u_i,t)$ from \eref{eq:Vzrelax} (the thick gray lines in \fref{fig:Vef}), the behavior of the adiabatic bins can be predicted: $\mathbf{P}^{\rm EF}(u_1,t)$ will not change direction because $V_{\nu\nu,z}^{\rm EF}(u_1,t)$ stays negative, and $\mathbf{P}^{\rm EF}(u_3,t)$ will change direction following  $V_{\nu\nu,z}^{\rm EF}(u_3,t)$. For the second bin, $\mathbf{P}^{\rm EF}(u_2,t)$ is expected to be non-adiabatic, so although $V_{\nu\nu,z}^{\rm EF}(u_2,t)$ changes direction at $t\sim 0$, $\mathbf{P}^{\rm EF}(u_2,t)$ cannot be expected to change direction as well. This is in agreement with the behavior we find for $C=3\;{\rm km}^{-1}$.
  For case B, all three bins are non-adiabatic for $C=0\;{\rm km}^{-1}$, and $\mathbf{P}^{\rm EF}(u_i,t)$ cannot be expected to follow $\mathbf{V}^{\rm EF}(u_i,t)$.
  Furthermore, it is clear from the right panel of \fref{fig:Vef} that with case B, the full solution for $V_{\nu\nu,z}^{\rm EF}(u_i,t)$ shows large deviations from the simple solution in \eref{eq:Vzrelax} indicated by thick gray lines. This is a non-adiabatic case without alignment; while we find that the conversion probability is suppressed in this case, in principle there could exist non-adiabatic cases with enhancement according to the interplay between flavor mixing, collisions and the shape of the angular distributions of neutrinos.
  Despite the simplifications intrinsic to this approach, the analysis gives qualitative understanding as well as quantitative predictions of fast flavor evolution in the  presence of collisions.

  To summarize, the overall behavior is that in case A an overall enhancement of flavor conversion is found. Such an enhancement is due to the combination of two effects due to collisions
and can be explained by invoking adiabaticity: Collisions gradually change the direction of $\mathbf{V}_{\nu\nu}^{\rm EF}$ in the eigenframe as shown in \fref{fig:Vef} and tend to isotropize the angular distributions. This implies that only some of the bins can flip their direction. The opposite situation arises for case B where the evolution is non-adiabatic; the initial conversion probability in the absence of conversions is very large, and due to the effect of collisions, the conversion probability is suppressed as the value of $C$ is increased.

%%%%%%%%%%%%%%%%%%%%%%%%%%%%%%%%%%%%%%%%%%%%%%%%%%%%%%%%%%%%%%%%%%%%%%%%%%%%%%%%%%%%%%%%%%%%%%%%
\section{Discussion}
\label{sec:discussion}

Our results are in good agreement with  existing literature on fast flavor conversion in the presence of collisions~\cite{Shalgar:2020wcx,Sasaki:2021zld,Martin:2021xyl}. We accommodate the apparently contradicting results showing enhancement due to collisions~\cite{Shalgar:2020wcx,Sasaki:2021zld} or suppression~\cite{Martin:2021xyl}. Different initial angular distributions result in different flavor outcomes due to the non-trivial interplay of  collisions and flavor conversion. Some cases allow for an enhancement (such as case A) while others only show suppression (such as case B). 

We empirically find that the enhancement of flavor conversion takes place as an effect of the isotropization of the angular distributions (because of collisions) when otherwise the angular distributions would have been such to induce little flavor conversion in the absence of collisions (e.g.~conversion probability $P_{ex} \lesssim 0.3$). 
When in the absence of collisions, the conversion probability would have been large (e.g.~conversion probability $P_{ex} \gtrsim 0.3$), then we would expect an overall suppression of flavor conversion resulting from the isotropization of the angular distributions as a result of collisions. Moreover, in case A, collisions sweep flavor conversion across the angular distribution gradually, while conversions remain localized in case B where flavor conversion is overall suppressed by collisions.

The detailed analysis we perform of the flavor conversion can be directly compared to that of Ref.~\cite{Sasaki:2021zld}. 
Their analysis focuses to a very large extent on the phase differences of $\mathbf{P}$ and $\mathbf{V}$, $\delta$. In place of going to an eigenframe, Ref.~\cite{Sasaki:2021zld} somewhat equivalently focuses on the alignment of the $x$ and $y$ components. In addition, it is found that the different angular bins evolve in a synchronized manner, which is also what we see when we transform to the eigenframe and all angular modes oscillate very slowly around the $z$ axis.
One key point where the approach of Ref.~\cite{Sasaki:2021zld} and ours differ is in the explanation of the enhancement of flavor conversion. Reference~\cite{Sasaki:2021zld}  focuses on an imbalance between the positive and negative values of $\delta$ induced by the collision term, which accumulates and leads to the enhancement. In this work, we find the enhancement is due to adiabatic evolution and the gradual relaxation of $V_{\nu\nu,z}^{\rm EF}$.

%%%%%%%%%%%%%%%%%%%%%%%%%%%%%%%%%%%%%%%%%%%%%%%%%%%%%%%%%%%%%%%%%%%%%%%%%%%%%%%%%%%%%%%%%%%%%%%%
\section{Conclusions}
\label{sec:conclusions}

In this paper, we investigate the apparent contradictory outcome of the interplay between fast flavor conversion and collisions  observed in Refs.~\cite{Shalgar:2020wcx,Sasaki:2021zld,Martin:2021xyl}. We  point out that direction-changing collisions may both enhance and suppress neutrino flavor conversion according to the initial angular distributions of the electron flavors. In order to do so, 
we have adopted two simple three bin neutrino models where the eigenvalues and eigenvectors of the linearized equations can be determined analytically; one (corresponding to almost uniform angular distributions) leading to an enhancement of flavor conversion, and the other one (with forward peaked angular distributions) showing suppression of flavor conversion in the presence of collisions. Our findings are, of course, limited in scope because of the approximations intrinsic to our model {\and a different choice of the angular distributions could lead to a different outcome; however  the mechanism of collisions gradually moving the region of flavor conversion across angular bins is  expected  to provide general insight into the non-trivial interplay between flavor conversion and collisions.

%%%%%%%%%%%%%%%%%%%%%%%%%%%%%%%%%%%%%%%%%%%%%%%%%%%%%%%%%%%%%%%%%%%%%%%%%%%%%%%%%%%%%%%%%%%%%%%%
\section*{Acknowledgements}
We thank Ian Padilla-Gay for useful discussions. This project has received funding from the Villum Foundation (Project No.~37358), the Danmarks Frie Forskningsfonds (Project No. 8049-00038B), the European Union's Horizon 2020 research and innovation program under the Marie Sklodowska-Curie grant agreement No.~847523 (``INTERACTIONS''), and the MERAC Foundation. 

%\appendix

\bibliography{literature}

\begin{thebibliography}{49}%
\makeatletter
\providecommand \@ifxundefined [1]{%
 \@ifx{#1\undefined}
}%
\providecommand \@ifnum [1]{%
 \ifnum #1\expandafter \@firstoftwo
 \else \expandafter \@secondoftwo
 \fi
}%
\providecommand \@ifx [1]{%
 \ifx #1\expandafter \@firstoftwo
 \else \expandafter \@secondoftwo
 \fi
}%
\providecommand \natexlab [1]{#1}%
\providecommand \enquote  [1]{``#1''}%
\providecommand \bibnamefont  [1]{#1}%
\providecommand \bibfnamefont [1]{#1}%
\providecommand \citenamefont [1]{#1}%
\providecommand \href@noop [0]{\@secondoftwo}%
\providecommand \href [0]{\begingroup \@sanitize@url \@href}%
\providecommand \@href[1]{\@@startlink{#1}\@@href}%
\providecommand \@@href[1]{\endgroup#1\@@endlink}%
\providecommand \@sanitize@url [0]{\catcode `\\12\catcode `\$12\catcode
  `\&12\catcode `\#12\catcode `\^12\catcode `\_12\catcode `\%12\relax}%
\providecommand \@@startlink[1]{}%
\providecommand \@@endlink[0]{}%
\providecommand \url  [0]{\begingroup\@sanitize@url \@url }%
\providecommand \@url [1]{\endgroup\@href {#1}{\urlprefix }}%
\providecommand \urlprefix  [0]{URL }%
\providecommand \Eprint [0]{\href }%
\providecommand \doibase [0]{http://dx.doi.org/}%
\providecommand \selectlanguage [0]{\@gobble}%
\providecommand \bibinfo  [0]{\@secondoftwo}%
\providecommand \bibfield  [0]{\@secondoftwo}%
\providecommand \translation [1]{[#1]}%
\providecommand \BibitemOpen [0]{}%
\providecommand \bibitemStop [0]{}%
\providecommand \bibitemNoStop [0]{.\EOS\space}%
\providecommand \EOS [0]{\spacefactor3000\relax}%
\providecommand \BibitemShut  [1]{\csname bibitem#1\endcsname}%
\let\auto@bib@innerbib\@empty
%</preamble>
\bibitem [{\citenamefont {Tamborra}\ and\ \citenamefont
  {Shalgar}(2021)}]{Tamborra:2020cul}%
  \BibitemOpen
  \bibfield  {author} {\bibinfo {author} {\bibfnamefont {Irene}\ \bibnamefont
  {Tamborra}}\ and\ \bibinfo {author} {\bibfnamefont {Shashank}\ \bibnamefont
  {Shalgar}},\ }\bibfield  {title} {\enquote {\bibinfo {title} {{New
  Developments in Flavor Evolution of a Dense Neutrino Gas}},}\ }\href
  {\doibase 10.1146/annurev-nucl-102920-050505} {\bibfield  {journal} {\bibinfo
   {journal} {Ann. Rev. Nucl. Part. Sci.}\ }\textbf {\bibinfo {volume} {71}},\
  \bibinfo {pages} {165--188} (\bibinfo {year} {2021})},\ \Eprint
  {http://arxiv.org/abs/2011.01948} {arXiv:2011.01948 [astro-ph.HE]}
  \BibitemShut {NoStop}%
\bibitem [{\citenamefont {Capozzi}\ and\ \citenamefont
  {Saviano}(2022)}]{Capozzi:2022slf}%
  \BibitemOpen
  \bibfield  {author} {\bibinfo {author} {\bibfnamefont {Francesco}\
  \bibnamefont {Capozzi}}\ and\ \bibinfo {author} {\bibfnamefont {Ninetta}\
  \bibnamefont {Saviano}},\ }\bibfield  {title} {\enquote {\bibinfo {title}
  {{Neutrino Flavor Conversions in High-Density Astrophysical and Cosmological
  Environments}},}\ }\href {\doibase 10.3390/universe8020094} {\bibfield
  {journal} {\bibinfo  {journal} {Universe}\ }\textbf {\bibinfo {volume} {8}},\
  \bibinfo {pages} {94} (\bibinfo {year} {2022})},\ \Eprint
  {http://arxiv.org/abs/2202.02494} {arXiv:2202.02494 [hep-ph]} \BibitemShut
  {NoStop}%
\bibitem [{\citenamefont {Mirizzi}\ \emph {et~al.}(2016)\citenamefont
  {Mirizzi}, \citenamefont {Tamborra}, \citenamefont {Janka}, \citenamefont
  {Saviano}, \citenamefont {Scholberg}, \citenamefont {Bollig}, \citenamefont
  {Hudepohl},\ and\ \citenamefont {Chakraborty}}]{Mirizzi:2015eza}%
  \BibitemOpen
  \bibfield  {author} {\bibinfo {author} {\bibfnamefont {Alessandro}\
  \bibnamefont {Mirizzi}}, \bibinfo {author} {\bibfnamefont {Irene}\
  \bibnamefont {Tamborra}}, \bibinfo {author} {\bibfnamefont {Hans-Thomas}\
  \bibnamefont {Janka}}, \bibinfo {author} {\bibfnamefont {Ninetta}\
  \bibnamefont {Saviano}}, \bibinfo {author} {\bibfnamefont {Kate}\
  \bibnamefont {Scholberg}}, \bibinfo {author} {\bibfnamefont {Robert}\
  \bibnamefont {Bollig}}, \bibinfo {author} {\bibfnamefont {Lorenz}\
  \bibnamefont {Hudepohl}}, \ and\ \bibinfo {author} {\bibfnamefont {Sovan}\
  \bibnamefont {Chakraborty}},\ }\bibfield  {title} {\enquote {\bibinfo {title}
  {{Supernova Neutrinos: Production, Oscillations and Detection}},}\ }\href
  {\doibase 10.1393/ncr/i2016-10120-8} {\bibfield  {journal} {\bibinfo
  {journal} {Riv. Nuovo Cim.}\ }\textbf {\bibinfo {volume} {39}},\ \bibinfo
  {pages} {1--112} (\bibinfo {year} {2016})},\ \Eprint
  {http://arxiv.org/abs/1508.00785} {arXiv:1508.00785 [astro-ph.HE]}
  \BibitemShut {NoStop}%
\bibitem [{\citenamefont {Duan}\ \emph {et~al.}(2010)\citenamefont {Duan},
  \citenamefont {Fuller},\ and\ \citenamefont {Qian}}]{Duan:2010bg}%
  \BibitemOpen
  \bibfield  {author} {\bibinfo {author} {\bibfnamefont {Huaiyu}\ \bibnamefont
  {Duan}}, \bibinfo {author} {\bibfnamefont {George~M.}\ \bibnamefont
  {Fuller}}, \ and\ \bibinfo {author} {\bibfnamefont {Yong-Zhong}\ \bibnamefont
  {Qian}},\ }\bibfield  {title} {\enquote {\bibinfo {title} {{Collective
  Neutrino Oscillations}},}\ }\href {\doibase
  10.1146/annurev.nucl.012809.104524} {\bibfield  {journal} {\bibinfo
  {journal} {Ann. Rev. Nucl. Part. Sci.}\ }\textbf {\bibinfo {volume} {60}},\
  \bibinfo {pages} {569--594} (\bibinfo {year} {2010})},\ \Eprint
  {http://arxiv.org/abs/1001.2799} {arXiv:1001.2799 [hep-ph]} \BibitemShut
  {NoStop}%
\bibitem [{\citenamefont {Stodolsky}(1987)}]{Stodolsky:1986dx}%
  \BibitemOpen
  \bibfield  {author} {\bibinfo {author} {\bibfnamefont {Leo}\ \bibnamefont
  {Stodolsky}},\ }\bibfield  {title} {\enquote {\bibinfo {title} {{On the
  Treatment of Neutrino Oscillations in a Thermal Environment}},}\ }\href
  {\doibase 10.1103/PhysRevD.36.2273} {\bibfield  {journal} {\bibinfo
  {journal} {Phys. Rev. D}\ }\textbf {\bibinfo {volume} {36}},\ \bibinfo
  {pages} {2273} (\bibinfo {year} {1987})}\BibitemShut {NoStop}%
\bibitem [{\citenamefont {Stodolsky}(1974)}]{Stodolsky:1974hm}%
  \BibitemOpen
  \bibfield  {author} {\bibinfo {author} {\bibfnamefont {Leo}\ \bibnamefont
  {Stodolsky}},\ }\bibfield  {title} {\enquote {\bibinfo {title} {{Neutron
  Optics and Weak Currents}},}\ }\href {\doibase 10.1016/0370-2693(74)90688-1}
  {\bibfield  {journal} {\bibinfo  {journal} {Phys. Lett. B}\ }\textbf
  {\bibinfo {volume} {50}},\ \bibinfo {pages} {352--356} (\bibinfo {year}
  {1974})}\BibitemShut {NoStop}%
\bibitem [{\citenamefont {Cherry}\ \emph {et~al.}(2012)\citenamefont {Cherry},
  \citenamefont {Carlson}, \citenamefont {Friedland}, \citenamefont {Fuller},\
  and\ \citenamefont {Vlasenko}}]{Cherry:2012zw}%
  \BibitemOpen
  \bibfield  {author} {\bibinfo {author} {\bibfnamefont {John~F.}\ \bibnamefont
  {Cherry}}, \bibinfo {author} {\bibfnamefont {J.}~\bibnamefont {Carlson}},
  \bibinfo {author} {\bibfnamefont {Alexander}\ \bibnamefont {Friedland}},
  \bibinfo {author} {\bibfnamefont {George~M.}\ \bibnamefont {Fuller}}, \ and\
  \bibinfo {author} {\bibfnamefont {Alexey}\ \bibnamefont {Vlasenko}},\
  }\bibfield  {title} {\enquote {\bibinfo {title} {{Neutrino scattering and
  flavor transformation in supernovae}},}\ }\href {\doibase
  10.1103/PhysRevLett.108.261104} {\bibfield  {journal} {\bibinfo  {journal}
  {Phys. Rev. Lett.}\ }\textbf {\bibinfo {volume} {108}},\ \bibinfo {pages}
  {261104} (\bibinfo {year} {2012})},\ \Eprint {http://arxiv.org/abs/1203.1607}
  {arXiv:1203.1607 [hep-ph]} \BibitemShut {NoStop}%
\bibitem [{\citenamefont {Sarikas}\ \emph {et~al.}(2012)\citenamefont
  {Sarikas}, \citenamefont {Tamborra}, \citenamefont {Raffelt}, \citenamefont
  {Hudepohl},\ and\ \citenamefont {Janka}}]{Sarikas:2012vb}%
  \BibitemOpen
  \bibfield  {author} {\bibinfo {author} {\bibfnamefont {Srdjan}\ \bibnamefont
  {Sarikas}}, \bibinfo {author} {\bibfnamefont {Irene}\ \bibnamefont
  {Tamborra}}, \bibinfo {author} {\bibfnamefont {Georg}\ \bibnamefont
  {Raffelt}}, \bibinfo {author} {\bibfnamefont {Lorenz}\ \bibnamefont
  {Hudepohl}}, \ and\ \bibinfo {author} {\bibfnamefont {Hans-Thomas}\
  \bibnamefont {Janka}},\ }\bibfield  {title} {\enquote {\bibinfo {title}
  {{Supernova neutrino halo and the suppression of self-induced flavor
  conversion}},}\ }\href {\doibase 10.1103/PhysRevD.85.113007} {\bibfield
  {journal} {\bibinfo  {journal} {Phys. Rev. D}\ }\textbf {\bibinfo {volume}
  {85}},\ \bibinfo {pages} {113007} (\bibinfo {year} {2012})},\ \Eprint
  {http://arxiv.org/abs/1204.0971} {arXiv:1204.0971 [hep-ph]} \BibitemShut
  {NoStop}%
\bibitem [{\citenamefont {Cherry}\ \emph {et~al.}(2013)\citenamefont {Cherry},
  \citenamefont {Carlson}, \citenamefont {Friedland}, \citenamefont {Fuller},\
  and\ \citenamefont {Vlasenko}}]{Cherry:2013mv}%
  \BibitemOpen
  \bibfield  {author} {\bibinfo {author} {\bibfnamefont {John~F.}\ \bibnamefont
  {Cherry}}, \bibinfo {author} {\bibfnamefont {J.}~\bibnamefont {Carlson}},
  \bibinfo {author} {\bibfnamefont {Alexander}\ \bibnamefont {Friedland}},
  \bibinfo {author} {\bibfnamefont {George~M.}\ \bibnamefont {Fuller}}, \ and\
  \bibinfo {author} {\bibfnamefont {Alexey}\ \bibnamefont {Vlasenko}},\
  }\bibfield  {title} {\enquote {\bibinfo {title} {{Halo Modification of a
  Supernova Neutronization Neutrino Burst}},}\ }\href {\doibase
  10.1103/PhysRevD.87.085037} {\bibfield  {journal} {\bibinfo  {journal} {Phys.
  Rev. D}\ }\textbf {\bibinfo {volume} {87}},\ \bibinfo {pages} {085037}
  (\bibinfo {year} {2013})},\ \Eprint {http://arxiv.org/abs/1302.1159}
  {arXiv:1302.1159 [astro-ph.HE]} \BibitemShut {NoStop}%
\bibitem [{\citenamefont {Cirigliano}\ \emph {et~al.}(2018)\citenamefont
  {Cirigliano}, \citenamefont {Paris},\ and\ \citenamefont
  {Shalgar}}]{Cirigliano:2018rst}%
  \BibitemOpen
  \bibfield  {author} {\bibinfo {author} {\bibfnamefont {Vincenzo}\
  \bibnamefont {Cirigliano}}, \bibinfo {author} {\bibfnamefont {Mark}\
  \bibnamefont {Paris}}, \ and\ \bibinfo {author} {\bibfnamefont {Shashank}\
  \bibnamefont {Shalgar}},\ }\bibfield  {title} {\enquote {\bibinfo {title}
  {{Collective neutrino oscillations with the halo effect in single-angle
  approximation}},}\ }\href {\doibase 10.1088/1475-7516/2018/11/019} {\bibfield
   {journal} {\bibinfo  {journal} {JCAP}\ }\textbf {\bibinfo {volume} {11}},\
  \bibinfo {pages} {019} (\bibinfo {year} {2018})},\ \Eprint
  {http://arxiv.org/abs/1807.07070} {arXiv:1807.07070 [hep-ph]} \BibitemShut
  {NoStop}%
\bibitem [{\citenamefont {Zaizen}\ \emph {et~al.}(2020)\citenamefont {Zaizen},
  \citenamefont {Cherry}, \citenamefont {Takiwaki}, \citenamefont {Horiuchi},
  \citenamefont {Kotake}, \citenamefont {Umeda},\ and\ \citenamefont
  {Yoshida}}]{Zaizen:2019ufj}%
  \BibitemOpen
  \bibfield  {author} {\bibinfo {author} {\bibfnamefont {Masamichi}\
  \bibnamefont {Zaizen}}, \bibinfo {author} {\bibfnamefont {John~F.}\
  \bibnamefont {Cherry}}, \bibinfo {author} {\bibfnamefont {Tomoya}\
  \bibnamefont {Takiwaki}}, \bibinfo {author} {\bibfnamefont {Shunsaku}\
  \bibnamefont {Horiuchi}}, \bibinfo {author} {\bibfnamefont {Kei}\
  \bibnamefont {Kotake}}, \bibinfo {author} {\bibfnamefont {Hideyuki}\
  \bibnamefont {Umeda}}, \ and\ \bibinfo {author} {\bibfnamefont {Takashi}\
  \bibnamefont {Yoshida}},\ }\bibfield  {title} {\enquote {\bibinfo {title}
  {{Neutrino halo effect on collective neutrino oscillation in iron
  core-collapse supernova model of a 9.6 $M_{\odot}$ star}},}\ }\href {\doibase
  10.1088/1475-7516/2020/06/011} {\bibfield  {journal} {\bibinfo  {journal}
  {JCAP}\ }\textbf {\bibinfo {volume} {06}},\ \bibinfo {pages} {011} (\bibinfo
  {year} {2020})},\ \Eprint {http://arxiv.org/abs/1908.10594} {arXiv:1908.10594
  [astro-ph.HE]} \BibitemShut {NoStop}%
\bibitem [{\citenamefont {Cherry}\ \emph {et~al.}(2020)\citenamefont {Cherry},
  \citenamefont {Fuller}, \citenamefont {Horiuchi}, \citenamefont {Kotake},
  \citenamefont {Takiwaki},\ and\ \citenamefont {Fischer}}]{Cherry:2019vkv}%
  \BibitemOpen
  \bibfield  {author} {\bibinfo {author} {\bibfnamefont {John~F.}\ \bibnamefont
  {Cherry}}, \bibinfo {author} {\bibfnamefont {George~M.}\ \bibnamefont
  {Fuller}}, \bibinfo {author} {\bibfnamefont {Shunsaku}\ \bibnamefont
  {Horiuchi}}, \bibinfo {author} {\bibfnamefont {Kei}\ \bibnamefont {Kotake}},
  \bibinfo {author} {\bibfnamefont {Tomoya}\ \bibnamefont {Takiwaki}}, \ and\
  \bibinfo {author} {\bibfnamefont {Tobias}\ \bibnamefont {Fischer}},\
  }\bibfield  {title} {\enquote {\bibinfo {title} {{Time of Flight and
  Supernova Progenitor Effects on the Neutrino Halo}},}\ }\href {\doibase
  10.1103/PhysRevD.102.023022} {\bibfield  {journal} {\bibinfo  {journal}
  {Phys. Rev. D}\ }\textbf {\bibinfo {volume} {102}},\ \bibinfo {pages}
  {023022} (\bibinfo {year} {2020})},\ \Eprint
  {http://arxiv.org/abs/1912.11489} {arXiv:1912.11489 [astro-ph.HE]}
  \BibitemShut {NoStop}%
\bibitem [{\citenamefont {Chakraborty}\ \emph
  {et~al.}(2016{\natexlab{a}})\citenamefont {Chakraborty}, \citenamefont
  {Hansen}, \citenamefont {Izaguirre},\ and\ \citenamefont
  {Raffelt}}]{Chakraborty:2016yeg}%
  \BibitemOpen
  \bibfield  {author} {\bibinfo {author} {\bibfnamefont {Sovan}\ \bibnamefont
  {Chakraborty}}, \bibinfo {author} {\bibfnamefont {Rasmus}\ \bibnamefont
  {Hansen}}, \bibinfo {author} {\bibfnamefont {Ignacio}\ \bibnamefont
  {Izaguirre}}, \ and\ \bibinfo {author} {\bibfnamefont {Georg}\ \bibnamefont
  {Raffelt}},\ }\bibfield  {title} {\enquote {\bibinfo {title} {{Collective
  neutrino flavor conversion: Recent developments}},}\ }\href {\doibase
  10.1016/j.nuclphysb.2016.02.012} {\bibfield  {journal} {\bibinfo  {journal}
  {Nucl. Phys. B}\ }\textbf {\bibinfo {volume} {908}},\ \bibinfo {pages}
  {366--381} (\bibinfo {year} {2016}{\natexlab{a}})},\ \Eprint
  {http://arxiv.org/abs/1602.02766} {arXiv:1602.02766 [hep-ph]} \BibitemShut
  {NoStop}%
\bibitem [{\citenamefont {Chakraborty}\ \emph
  {et~al.}(2016{\natexlab{b}})\citenamefont {Chakraborty}, \citenamefont
  {Hansen}, \citenamefont {Izaguirre},\ and\ \citenamefont
  {Raffelt}}]{Chakraborty:2016lct}%
  \BibitemOpen
  \bibfield  {author} {\bibinfo {author} {\bibfnamefont {Sovan}\ \bibnamefont
  {Chakraborty}}, \bibinfo {author} {\bibfnamefont {Rasmus~Sloth}\ \bibnamefont
  {Hansen}}, \bibinfo {author} {\bibfnamefont {Ignacio}\ \bibnamefont
  {Izaguirre}}, \ and\ \bibinfo {author} {\bibfnamefont {Georg}\ \bibnamefont
  {Raffelt}},\ }\bibfield  {title} {\enquote {\bibinfo {title} {{Self-induced
  neutrino flavor conversion without flavor mixing}},}\ }\href {\doibase
  10.1088/1475-7516/2016/03/042} {\bibfield  {journal} {\bibinfo  {journal}
  {JCAP}\ }\textbf {\bibinfo {volume} {03}},\ \bibinfo {pages} {042} (\bibinfo
  {year} {2016}{\natexlab{b}})},\ \Eprint {http://arxiv.org/abs/1602.00698}
  {arXiv:1602.00698 [hep-ph]} \BibitemShut {NoStop}%
\bibitem [{\citenamefont {Izaguirre}\ \emph {et~al.}(2017)\citenamefont
  {Izaguirre}, \citenamefont {Raffelt},\ and\ \citenamefont
  {Tamborra}}]{Izaguirre:2016gsx}%
  \BibitemOpen
  \bibfield  {author} {\bibinfo {author} {\bibfnamefont {Ignacio}\ \bibnamefont
  {Izaguirre}}, \bibinfo {author} {\bibfnamefont {Georg}\ \bibnamefont
  {Raffelt}}, \ and\ \bibinfo {author} {\bibfnamefont {Irene}\ \bibnamefont
  {Tamborra}},\ }\bibfield  {title} {\enquote {\bibinfo {title} {{Fast Pairwise
  Conversion of Supernova Neutrinos: A Dispersion-Relation Approach}},}\ }\href
  {\doibase 10.1103/PhysRevLett.118.021101} {\bibfield  {journal} {\bibinfo
  {journal} {Phys. Rev. Lett.}\ }\textbf {\bibinfo {volume} {118}},\ \bibinfo
  {pages} {021101} (\bibinfo {year} {2017})},\ \Eprint
  {http://arxiv.org/abs/1610.01612} {arXiv:1610.01612 [hep-ph]} \BibitemShut
  {NoStop}%
\bibitem [{\citenamefont {Morinaga}(2021)}]{Morinaga:2021vmc}%
  \BibitemOpen
  \bibfield  {author} {\bibinfo {author} {\bibfnamefont {Taiki}\ \bibnamefont
  {Morinaga}},\ }\bibfield  {title} {\enquote {\bibinfo {title} {{Fast neutrino
  flavor instability and neutrino flavor lepton number crossings}},}\
  }\href@noop {} {\  (\bibinfo {year} {2021})},\ \Eprint
  {http://arxiv.org/abs/2103.15267} {arXiv:2103.15267 [hep-ph]} \BibitemShut
  {NoStop}%
\bibitem [{\citenamefont {Dasgupta}(2022)}]{Dasgupta:2021gfs}%
  \BibitemOpen
  \bibfield  {author} {\bibinfo {author} {\bibfnamefont {Basudeb}\ \bibnamefont
  {Dasgupta}},\ }\bibfield  {title} {\enquote {\bibinfo {title} {{Collective
  Neutrino Flavor Instability Requires a Crossing}},}\ }\href {\doibase
  10.1103/PhysRevLett.128.081102} {\bibfield  {journal} {\bibinfo  {journal}
  {Phys. Rev. Lett.}\ }\textbf {\bibinfo {volume} {128}},\ \bibinfo {pages}
  {081102} (\bibinfo {year} {2022})},\ \Eprint
  {http://arxiv.org/abs/2110.00192} {arXiv:2110.00192 [hep-ph]} \BibitemShut
  {NoStop}%
\bibitem [{\citenamefont {Shalgar}\ and\ \citenamefont
  {Tamborra}(2019)}]{Shalgar:2019kzy}%
  \BibitemOpen
  \bibfield  {author} {\bibinfo {author} {\bibfnamefont {Shashank}\
  \bibnamefont {Shalgar}}\ and\ \bibinfo {author} {\bibfnamefont {Irene}\
  \bibnamefont {Tamborra}},\ }\bibfield  {title} {\enquote {\bibinfo {title}
  {{On the Occurrence of Crossings Between the Angular Distributions of
  Electron Neutrinos and Antineutrinos in the Supernova Core}},}\ }\href
  {\doibase 10.3847/1538-4357/ab38ba} {\bibfield  {journal} {\bibinfo
  {journal} {Astrophys. J.}\ }\textbf {\bibinfo {volume} {883}},\ \bibinfo
  {pages} {80} (\bibinfo {year} {2019})},\ \Eprint
  {http://arxiv.org/abs/1904.07236} {arXiv:1904.07236 [astro-ph.HE]}
  \BibitemShut {NoStop}%
\bibitem [{\citenamefont {Morinaga}\ \emph {et~al.}(2020)\citenamefont
  {Morinaga}, \citenamefont {Nagakura}, \citenamefont {Kato},\ and\
  \citenamefont {Yamada}}]{Morinaga:2019wsv}%
  \BibitemOpen
  \bibfield  {author} {\bibinfo {author} {\bibfnamefont {Taiki}\ \bibnamefont
  {Morinaga}}, \bibinfo {author} {\bibfnamefont {Hiroki}\ \bibnamefont
  {Nagakura}}, \bibinfo {author} {\bibfnamefont {Chinami}\ \bibnamefont
  {Kato}}, \ and\ \bibinfo {author} {\bibfnamefont {Shoichi}\ \bibnamefont
  {Yamada}},\ }\bibfield  {title} {\enquote {\bibinfo {title} {{Fast
  neutrino-flavor conversion in the preshock region of core-collapse
  supernovae}},}\ }\href {\doibase 10.1103/PhysRevResearch.2.012046} {\bibfield
   {journal} {\bibinfo  {journal} {Phys. Rev. Res.}\ }\textbf {\bibinfo
  {volume} {2}},\ \bibinfo {pages} {012046(R)} (\bibinfo {year} {2020})},\
  \Eprint {http://arxiv.org/abs/1909.13131} {arXiv:1909.13131 [astro-ph.HE]}
  \BibitemShut {NoStop}%
\bibitem [{\citenamefont {Nagakura}\ \emph {et~al.}(2019)\citenamefont
  {Nagakura}, \citenamefont {Morinaga}, \citenamefont {Kato},\ and\
  \citenamefont {Yamada}}]{Nagakura:2019sig}%
  \BibitemOpen
  \bibfield  {author} {\bibinfo {author} {\bibfnamefont {Hiroki}\ \bibnamefont
  {Nagakura}}, \bibinfo {author} {\bibfnamefont {Taiki}\ \bibnamefont
  {Morinaga}}, \bibinfo {author} {\bibfnamefont {Chinami}\ \bibnamefont
  {Kato}}, \ and\ \bibinfo {author} {\bibfnamefont {Shoichi}\ \bibnamefont
  {Yamada}},\ }\bibfield  {title} {\enquote {\bibinfo {title} {{Fast-pairwise
  collective neutrino oscillations associated with asymmetric neutrino
  emissions in core-collapse supernova}},}\ }\href@noop {} {\  (\bibinfo {year}
  {2019})},\ \Eprint {http://arxiv.org/abs/1910.04288} {arXiv:1910.04288
  [astro-ph.HE]} \BibitemShut {NoStop}%
\bibitem [{\citenamefont {Glas}\ \emph {et~al.}(2020)\citenamefont {Glas},
  \citenamefont {Janka}, \citenamefont {Capozzi}, \citenamefont {Sen},
  \citenamefont {Dasgupta}, \citenamefont {Mirizzi},\ and\ \citenamefont
  {Sigl}}]{Glas:2019ijo}%
  \BibitemOpen
  \bibfield  {author} {\bibinfo {author} {\bibfnamefont {Robert}\ \bibnamefont
  {Glas}}, \bibinfo {author} {\bibfnamefont {H.Thomas}\ \bibnamefont {Janka}},
  \bibinfo {author} {\bibfnamefont {Francesco}\ \bibnamefont {Capozzi}},
  \bibinfo {author} {\bibfnamefont {Manibrata}\ \bibnamefont {Sen}}, \bibinfo
  {author} {\bibfnamefont {Basudeb}\ \bibnamefont {Dasgupta}}, \bibinfo
  {author} {\bibfnamefont {Alessandro}\ \bibnamefont {Mirizzi}}, \ and\
  \bibinfo {author} {\bibfnamefont {Guenter}\ \bibnamefont {Sigl}},\ }\bibfield
   {title} {\enquote {\bibinfo {title} {{Fast Neutrino Flavor Instability in
  the Neutron-star Convection Layer of Three-dimensional Supernova Models}},}\
  }\href {\doibase 10.1103/PhysRevD.101.063001} {\bibfield  {journal} {\bibinfo
   {journal} {Phys. Rev. D}\ }\textbf {\bibinfo {volume} {101}},\ \bibinfo
  {pages} {063001} (\bibinfo {year} {2020})},\ \Eprint
  {http://arxiv.org/abs/1912.00274} {arXiv:1912.00274 [astro-ph.HE]}
  \BibitemShut {NoStop}%
\bibitem [{\citenamefont {Abbar}\ \emph {et~al.}(2020)\citenamefont {Abbar},
  \citenamefont {Duan}, \citenamefont {Sumiyoshi}, \citenamefont {Takiwaki},\
  and\ \citenamefont {Volpe}}]{Abbar:2019zoq}%
  \BibitemOpen
  \bibfield  {author} {\bibinfo {author} {\bibfnamefont {Sajad}\ \bibnamefont
  {Abbar}}, \bibinfo {author} {\bibfnamefont {Huaiyu}\ \bibnamefont {Duan}},
  \bibinfo {author} {\bibfnamefont {Kohsuke}\ \bibnamefont {Sumiyoshi}},
  \bibinfo {author} {\bibfnamefont {Tomoya}\ \bibnamefont {Takiwaki}}, \ and\
  \bibinfo {author} {\bibfnamefont {Maria~Cristina}\ \bibnamefont {Volpe}},\
  }\bibfield  {title} {\enquote {\bibinfo {title} {{Fast Neutrino Flavor
  Conversion Modes in Multidimensional Core-collapse Supernova Models: the Role
  of the Asymmetric Neutrino Distributions}},}\ }\href {\doibase
  10.1103/PhysRevD.101.043016} {\bibfield  {journal} {\bibinfo  {journal}
  {Phys. Rev. D}\ }\textbf {\bibinfo {volume} {101}},\ \bibinfo {pages}
  {043016} (\bibinfo {year} {2020})},\ \Eprint
  {http://arxiv.org/abs/1911.01983} {arXiv:1911.01983 [astro-ph.HE]}
  \BibitemShut {NoStop}%
\bibitem [{\citenamefont {Abbar}\ \emph {et~al.}(2021)\citenamefont {Abbar},
  \citenamefont {Capozzi}, \citenamefont {Glas}, \citenamefont {Janka},\ and\
  \citenamefont {Tamborra}}]{Abbar:2020qpi}%
  \BibitemOpen
  \bibfield  {author} {\bibinfo {author} {\bibfnamefont {Sajad}\ \bibnamefont
  {Abbar}}, \bibinfo {author} {\bibfnamefont {Francesco}\ \bibnamefont
  {Capozzi}}, \bibinfo {author} {\bibfnamefont {Robert}\ \bibnamefont {Glas}},
  \bibinfo {author} {\bibfnamefont {H.Thomas}\ \bibnamefont {Janka}}, \ and\
  \bibinfo {author} {\bibfnamefont {Irene}\ \bibnamefont {Tamborra}},\
  }\bibfield  {title} {\enquote {\bibinfo {title} {{On the characteristics of
  fast neutrino flavor instabilities in three-dimensional core-collapse
  supernova models}},}\ }\href {\doibase 10.1103/PhysRevD.103.063033}
  {\bibfield  {journal} {\bibinfo  {journal} {Phys. Rev. D}\ }\textbf {\bibinfo
  {volume} {103}},\ \bibinfo {pages} {063033} (\bibinfo {year} {2021})},\
  \Eprint {http://arxiv.org/abs/2012.06594} {arXiv:2012.06594 [astro-ph.HE]}
  \BibitemShut {NoStop}%
\bibitem [{\citenamefont {Nagakura}\ \emph {et~al.}(2021)\citenamefont
  {Nagakura}, \citenamefont {Burrows}, \citenamefont {Johns},\ and\
  \citenamefont {Fuller}}]{Nagakura:2021hyb}%
  \BibitemOpen
  \bibfield  {author} {\bibinfo {author} {\bibfnamefont {Hiroki}\ \bibnamefont
  {Nagakura}}, \bibinfo {author} {\bibfnamefont {Adam}\ \bibnamefont
  {Burrows}}, \bibinfo {author} {\bibfnamefont {Lucas}\ \bibnamefont {Johns}},
  \ and\ \bibinfo {author} {\bibfnamefont {George~M.}\ \bibnamefont {Fuller}},\
  }\bibfield  {title} {\enquote {\bibinfo {title} {{Where, when, and why:
  Occurrence of fast-pairwise collective neutrino oscillation in
  three-dimensional core-collapse supernova models}},}\ }\href {\doibase
  10.1103/PhysRevD.104.083025} {\bibfield  {journal} {\bibinfo  {journal}
  {Phys. Rev. D}\ }\textbf {\bibinfo {volume} {104}},\ \bibinfo {pages}
  {083025} (\bibinfo {year} {2021})},\ \Eprint
  {http://arxiv.org/abs/2108.07281} {arXiv:2108.07281 [astro-ph.HE]}
  \BibitemShut {NoStop}%
\bibitem [{\citenamefont {Brandt}\ \emph {et~al.}(2011)\citenamefont {Brandt},
  \citenamefont {Burrows}, \citenamefont {Ott},\ and\ \citenamefont
  {Livne}}]{Brandt:2010xa}%
  \BibitemOpen
  \bibfield  {author} {\bibinfo {author} {\bibfnamefont {Timothy~D.}\
  \bibnamefont {Brandt}}, \bibinfo {author} {\bibfnamefont {Adam}\ \bibnamefont
  {Burrows}}, \bibinfo {author} {\bibfnamefont {Christian~D.}\ \bibnamefont
  {Ott}}, \ and\ \bibinfo {author} {\bibfnamefont {Eli}\ \bibnamefont
  {Livne}},\ }\bibfield  {title} {\enquote {\bibinfo {title} {{Results From
  Core-Collapse Simulations with Multi-Dimensional, Multi-Angle Neutrino
  Transport}},}\ }\href {\doibase 10.1088/0004-637X/728/1/8} {\bibfield
  {journal} {\bibinfo  {journal} {Astrophys. J.}\ }\textbf {\bibinfo {volume}
  {728}},\ \bibinfo {pages} {8} (\bibinfo {year} {2011})},\ \Eprint
  {http://arxiv.org/abs/1009.4654} {arXiv:1009.4654 [astro-ph.HE]} \BibitemShut
  {NoStop}%
\bibitem [{\citenamefont {Johns}(2021)}]{Johns:2021qby}%
  \BibitemOpen
  \bibfield  {author} {\bibinfo {author} {\bibfnamefont {Lucas}\ \bibnamefont
  {Johns}},\ }\bibfield  {title} {\enquote {\bibinfo {title} {{Collisional
  flavor instabilities of supernova neutrinos}},}\ }\href@noop {} {\  (\bibinfo
  {year} {2021})},\ \Eprint {http://arxiv.org/abs/2104.11369} {arXiv:2104.11369
  [hep-ph]} \BibitemShut {NoStop}%
\bibitem [{\citenamefont {Sigl}(2022)}]{Sigl:2021tmj}%
  \BibitemOpen
  \bibfield  {author} {\bibinfo {author} {\bibfnamefont {G\"unter}\
  \bibnamefont {Sigl}},\ }\bibfield  {title} {\enquote {\bibinfo {title}
  {{Simulations of fast neutrino flavor conversions with interactions in
  inhomogeneous media}},}\ }\href {\doibase 10.1103/PhysRevD.105.043005}
  {\bibfield  {journal} {\bibinfo  {journal} {Phys. Rev. D}\ }\textbf {\bibinfo
  {volume} {105}},\ \bibinfo {pages} {043005} (\bibinfo {year} {2022})},\
  \Eprint {http://arxiv.org/abs/2109.00091} {arXiv:2109.00091 [hep-ph]}
  \BibitemShut {NoStop}%
\bibitem [{\citenamefont {Capozzi}\ \emph {et~al.}(2019)\citenamefont
  {Capozzi}, \citenamefont {Dasgupta}, \citenamefont {Mirizzi}, \citenamefont
  {Sen},\ and\ \citenamefont {Sigl}}]{Capozzi:2018clo}%
  \BibitemOpen
  \bibfield  {author} {\bibinfo {author} {\bibfnamefont {Francesco}\
  \bibnamefont {Capozzi}}, \bibinfo {author} {\bibfnamefont {Basudeb}\
  \bibnamefont {Dasgupta}}, \bibinfo {author} {\bibfnamefont {Alessandro}\
  \bibnamefont {Mirizzi}}, \bibinfo {author} {\bibfnamefont {Manibrata}\
  \bibnamefont {Sen}}, \ and\ \bibinfo {author} {\bibfnamefont {G\"unter}\
  \bibnamefont {Sigl}},\ }\bibfield  {title} {\enquote {\bibinfo {title}
  {{Collisional triggering of fast flavor conversions of supernova
  neutrinos}},}\ }\href {\doibase 10.1103/PhysRevLett.122.091101} {\bibfield
  {journal} {\bibinfo  {journal} {Phys. Rev. Lett.}\ }\textbf {\bibinfo
  {volume} {122}},\ \bibinfo {pages} {091101} (\bibinfo {year} {2019})},\
  \Eprint {http://arxiv.org/abs/1808.06618} {arXiv:1808.06618 [hep-ph]}
  \BibitemShut {NoStop}%
\bibitem [{\citenamefont {Shalgar}\ and\ \citenamefont
  {Tamborra}(2021{\natexlab{a}})}]{Shalgar:2020wcx}%
  \BibitemOpen
  \bibfield  {author} {\bibinfo {author} {\bibfnamefont {Shashank}\
  \bibnamefont {Shalgar}}\ and\ \bibinfo {author} {\bibfnamefont {Irene}\
  \bibnamefont {Tamborra}},\ }\bibfield  {title} {\enquote {\bibinfo {title}
  {{A change of direction in pairwise neutrino conversion physics: The effect
  of collisions}},}\ }\href {\doibase 10.1103/PhysRevD.103.063002} {\bibfield
  {journal} {\bibinfo  {journal} {Phys. Rev. D}\ }\textbf {\bibinfo {volume}
  {103}},\ \bibinfo {pages} {063002} (\bibinfo {year} {2021}{\natexlab{a}})},\
  \Eprint {http://arxiv.org/abs/2011.00004} {arXiv:2011.00004 [astro-ph.HE]}
  \BibitemShut {NoStop}%
\bibitem [{\citenamefont {Sasaki}\ and\ \citenamefont
  {Takiwaki}(2021)}]{Sasaki:2021zld}%
  \BibitemOpen
  \bibfield  {author} {\bibinfo {author} {\bibfnamefont {Hirokazu}\
  \bibnamefont {Sasaki}}\ and\ \bibinfo {author} {\bibfnamefont {Tomoya}\
  \bibnamefont {Takiwaki}},\ }\bibfield  {title} {\enquote {\bibinfo {title}
  {{Dynamics of fast neutrino flavor conversions with scattering effects: a
  detailed analysis}},}\ }\href@noop {} {\  (\bibinfo {year} {2021})},\ \Eprint
  {http://arxiv.org/abs/2109.14011} {arXiv:2109.14011 [hep-ph]} \BibitemShut
  {NoStop}%
\bibitem [{\citenamefont {Martin}\ \emph {et~al.}(2021)\citenamefont {Martin},
  \citenamefont {Carlson}, \citenamefont {Cirigliano},\ and\ \citenamefont
  {Duan}}]{Martin:2021xyl}%
  \BibitemOpen
  \bibfield  {author} {\bibinfo {author} {\bibfnamefont {Joshua~D.}\
  \bibnamefont {Martin}}, \bibinfo {author} {\bibfnamefont {J.}~\bibnamefont
  {Carlson}}, \bibinfo {author} {\bibfnamefont {Vincenzo}\ \bibnamefont
  {Cirigliano}}, \ and\ \bibinfo {author} {\bibfnamefont {Huaiyu}\ \bibnamefont
  {Duan}},\ }\bibfield  {title} {\enquote {\bibinfo {title} {{Fast flavor
  oscillations in dense neutrino media with collisions}},}\ }\href {\doibase
  10.1103/PhysRevD.103.063001} {\bibfield  {journal} {\bibinfo  {journal}
  {Phys. Rev. D}\ }\textbf {\bibinfo {volume} {103}},\ \bibinfo {pages}
  {063001} (\bibinfo {year} {2021})},\ \Eprint
  {http://arxiv.org/abs/2101.01278} {arXiv:2101.01278 [hep-ph]} \BibitemShut
  {NoStop}%
\bibitem [{\citenamefont {Banerjee}\ \emph {et~al.}(2011)\citenamefont
  {Banerjee}, \citenamefont {Dighe},\ and\ \citenamefont
  {Raffelt}}]{Banerjee:2011fj}%
  \BibitemOpen
  \bibfield  {author} {\bibinfo {author} {\bibfnamefont {Arka}\ \bibnamefont
  {Banerjee}}, \bibinfo {author} {\bibfnamefont {Amol}\ \bibnamefont {Dighe}},
  \ and\ \bibinfo {author} {\bibfnamefont {Georg}\ \bibnamefont {Raffelt}},\
  }\bibfield  {title} {\enquote {\bibinfo {title} {{Linearized flavor-stability
  analysis of dense neutrino streams}},}\ }\href {\doibase
  10.1103/PhysRevD.84.053013} {\bibfield  {journal} {\bibinfo  {journal} {Phys.
  Rev. D}\ }\textbf {\bibinfo {volume} {84}},\ \bibinfo {pages} {053013}
  (\bibinfo {year} {2011})},\ \Eprint {http://arxiv.org/abs/1107.2308}
  {arXiv:1107.2308 [hep-ph]} \BibitemShut {NoStop}%
\bibitem [{\citenamefont {Padilla-Gay}\ \emph {et~al.}(2022)\citenamefont
  {Padilla-Gay}, \citenamefont {Tamborra},\ and\ \citenamefont
  {Raffelt}}]{Padilla-Gay:2021haz}%
  \BibitemOpen
  \bibfield  {author} {\bibinfo {author} {\bibfnamefont {Ian}\ \bibnamefont
  {Padilla-Gay}}, \bibinfo {author} {\bibfnamefont {Irene}\ \bibnamefont
  {Tamborra}}, \ and\ \bibinfo {author} {\bibfnamefont {Georg~G.}\ \bibnamefont
  {Raffelt}},\ }\bibfield  {title} {\enquote {\bibinfo {title} {{Neutrino
  Flavor Pendulum Reloaded: The Case of Fast Pairwise Conversion}},}\ }\href
  {\doibase 10.1103/PhysRevLett.128.121102} {\bibfield  {journal} {\bibinfo
  {journal} {Phys. Rev. Lett.}\ }\textbf {\bibinfo {volume} {128}},\ \bibinfo
  {pages} {121102} (\bibinfo {year} {2022})},\ \Eprint
  {http://arxiv.org/abs/2109.14627} {arXiv:2109.14627 [astro-ph.HE]}
  \BibitemShut {NoStop}%
\bibitem [{\citenamefont {Shalgar}\ and\ \citenamefont
  {Tamborra}(2021{\natexlab{b}})}]{Shalgar:2021wlj}%
  \BibitemOpen
  \bibfield  {author} {\bibinfo {author} {\bibfnamefont {Shashank}\
  \bibnamefont {Shalgar}}\ and\ \bibinfo {author} {\bibfnamefont {Irene}\
  \bibnamefont {Tamborra}},\ }\bibfield  {title} {\enquote {\bibinfo {title}
  {{Three flavor revolution in fast pairwise neutrino conversion}},}\ }\href
  {\doibase 10.1103/PhysRevD.104.023011} {\bibfield  {journal} {\bibinfo
  {journal} {Phys. Rev. D}\ }\textbf {\bibinfo {volume} {104}},\ \bibinfo
  {pages} {023011} (\bibinfo {year} {2021}{\natexlab{b}})},\ \Eprint
  {http://arxiv.org/abs/2103.12743} {arXiv:2103.12743 [hep-ph]} \BibitemShut
  {NoStop}%
\bibitem [{\citenamefont {Capozzi}\ \emph {et~al.}(2020)\citenamefont
  {Capozzi}, \citenamefont {Chakraborty}, \citenamefont {Chakraborty},\ and\
  \citenamefont {Sen}}]{Capozzi:2020kge}%
  \BibitemOpen
  \bibfield  {author} {\bibinfo {author} {\bibfnamefont {Francesco}\
  \bibnamefont {Capozzi}}, \bibinfo {author} {\bibfnamefont {Madhurima}\
  \bibnamefont {Chakraborty}}, \bibinfo {author} {\bibfnamefont {Sovan}\
  \bibnamefont {Chakraborty}}, \ and\ \bibinfo {author} {\bibfnamefont
  {Manibrata}\ \bibnamefont {Sen}},\ }\bibfield  {title} {\enquote {\bibinfo
  {title} {{Fast flavor conversions in supernovae: the rise of mu-tau
  neutrinos}},}\ }\href {\doibase 10.1103/PhysRevLett.125.251801} {\bibfield
  {journal} {\bibinfo  {journal} {Phys. Rev. Lett.}\ }\textbf {\bibinfo
  {volume} {125}},\ \bibinfo {pages} {251801} (\bibinfo {year} {2020})},\
  \Eprint {http://arxiv.org/abs/2005.14204} {arXiv:2005.14204 [hep-ph]}
  \BibitemShut {NoStop}%
\bibitem [{\citenamefont {Chakraborty}\ and\ \citenamefont
  {Chakraborty}(2020)}]{Chakraborty:2019wxe}%
  \BibitemOpen
  \bibfield  {author} {\bibinfo {author} {\bibfnamefont {Madhurima}\
  \bibnamefont {Chakraborty}}\ and\ \bibinfo {author} {\bibfnamefont {Sovan}\
  \bibnamefont {Chakraborty}},\ }\bibfield  {title} {\enquote {\bibinfo {title}
  {{Three flavor neutrino conversions in supernovae: slow \& fast
  instabilities}},}\ }\href {\doibase 10.1088/1475-7516/2020/01/005} {\bibfield
   {journal} {\bibinfo  {journal} {JCAP}\ }\textbf {\bibinfo {volume} {01}},\
  \bibinfo {pages} {005} (\bibinfo {year} {2020})},\ \Eprint
  {http://arxiv.org/abs/1909.10420} {arXiv:1909.10420 [hep-ph]} \BibitemShut
  {NoStop}%
\bibitem [{\citenamefont {Sigl}\ and\ \citenamefont
  {Raffelt}(1993)}]{Sigl:1992fn}%
  \BibitemOpen
  \bibfield  {author} {\bibinfo {author} {\bibfnamefont {G.}~\bibnamefont
  {Sigl}}\ and\ \bibinfo {author} {\bibfnamefont {G.}~\bibnamefont {Raffelt}},\
  }\bibfield  {title} {\enquote {\bibinfo {title} {{General kinetic description
  of relativistic mixed neutrinos}},}\ }\href {\doibase
  10.1016/0550-3213(93)90175-O} {\bibfield  {journal} {\bibinfo  {journal}
  {Nucl. Phys. B}\ }\textbf {\bibinfo {volume} {406}},\ \bibinfo {pages}
  {423--451} (\bibinfo {year} {1993})}\BibitemShut {NoStop}%
\bibitem [{\citenamefont {Vlasenko}\ \emph {et~al.}(2014)\citenamefont
  {Vlasenko}, \citenamefont {Fuller},\ and\ \citenamefont
  {Cirigliano}}]{Vlasenko:2013fja}%
  \BibitemOpen
  \bibfield  {author} {\bibinfo {author} {\bibfnamefont {Alexey}\ \bibnamefont
  {Vlasenko}}, \bibinfo {author} {\bibfnamefont {George~M.}\ \bibnamefont
  {Fuller}}, \ and\ \bibinfo {author} {\bibfnamefont {Vincenzo}\ \bibnamefont
  {Cirigliano}},\ }\bibfield  {title} {\enquote {\bibinfo {title} {{Neutrino
  Quantum Kinetics}},}\ }\href {\doibase 10.1103/PhysRevD.89.105004} {\bibfield
   {journal} {\bibinfo  {journal} {Phys. Rev. D}\ }\textbf {\bibinfo {volume}
  {89}},\ \bibinfo {pages} {105004} (\bibinfo {year} {2014})},\ \Eprint
  {http://arxiv.org/abs/1309.2628} {arXiv:1309.2628 [hep-ph]} \BibitemShut
  {NoStop}%
\bibitem [{\citenamefont {Volpe}\ \emph {et~al.}(2013)\citenamefont {Volpe},
  \citenamefont {V\"a\"an\"anen},\ and\ \citenamefont
  {Espinoza}}]{Volpe:2013uxl}%
  \BibitemOpen
  \bibfield  {author} {\bibinfo {author} {\bibfnamefont {Cristina}\
  \bibnamefont {Volpe}}, \bibinfo {author} {\bibfnamefont {Daavid}\
  \bibnamefont {V\"a\"an\"anen}}, \ and\ \bibinfo {author} {\bibfnamefont
  {Catalina}\ \bibnamefont {Espinoza}},\ }\bibfield  {title} {\enquote
  {\bibinfo {title} {{Extended evolution equations for neutrino propagation in
  astrophysical and cosmological environments}},}\ }\href {\doibase
  10.1103/PhysRevD.87.113010} {\bibfield  {journal} {\bibinfo  {journal} {Phys.
  Rev. D}\ }\textbf {\bibinfo {volume} {87}},\ \bibinfo {pages} {113010}
  (\bibinfo {year} {2013})},\ \Eprint {http://arxiv.org/abs/1302.2374}
  {arXiv:1302.2374 [hep-ph]} \BibitemShut {NoStop}%
\bibitem [{\citenamefont {Shalgar}\ and\ \citenamefont
  {Tamborra}(2021{\natexlab{c}})}]{Shalgar:2020xns}%
  \BibitemOpen
  \bibfield  {author} {\bibinfo {author} {\bibfnamefont {Shashank}\
  \bibnamefont {Shalgar}}\ and\ \bibinfo {author} {\bibfnamefont {Irene}\
  \bibnamefont {Tamborra}},\ }\bibfield  {title} {\enquote {\bibinfo {title}
  {{Dispelling a myth on dense neutrino media: fast pairwise conversions depend
  on energy}},}\ }\href {\doibase 10.1088/1475-7516/2021/01/014} {\bibfield
  {journal} {\bibinfo  {journal} {JCAP}\ }\textbf {\bibinfo {volume} {01}},\
  \bibinfo {pages} {014} (\bibinfo {year} {2021}{\natexlab{c}})},\ \Eprint
  {http://arxiv.org/abs/2007.07926} {arXiv:2007.07926 [astro-ph.HE]}
  \BibitemShut {NoStop}%
\bibitem [{\citenamefont {Shalgar}\ and\ \citenamefont
  {Tamborra}(2022)}]{Shalgar:2021oko}%
  \BibitemOpen
  \bibfield  {author} {\bibinfo {author} {\bibfnamefont {Shashank}\
  \bibnamefont {Shalgar}}\ and\ \bibinfo {author} {\bibfnamefont {Irene}\
  \bibnamefont {Tamborra}},\ }\bibfield  {title} {\enquote {\bibinfo {title}
  {{Symmetry breaking induced by pairwise conversion of neutrinos in compact
  sources}},}\ }\href {\doibase 10.1103/PhysRevD.105.043018} {\bibfield
  {journal} {\bibinfo  {journal} {Phys. Rev. D}\ }\textbf {\bibinfo {volume}
  {105}},\ \bibinfo {pages} {043018} (\bibinfo {year} {2022})},\ \Eprint
  {http://arxiv.org/abs/2106.15622} {arXiv:2106.15622 [hep-ph]} \BibitemShut
  {NoStop}%
\bibitem [{\citenamefont {Shalgar}\ \emph {et~al.}(2020)\citenamefont
  {Shalgar}, \citenamefont {Padilla-Gay},\ and\ \citenamefont
  {Tamborra}}]{Shalgar:2019qwg}%
  \BibitemOpen
  \bibfield  {author} {\bibinfo {author} {\bibfnamefont {Shashank}\
  \bibnamefont {Shalgar}}, \bibinfo {author} {\bibfnamefont {Ian}\ \bibnamefont
  {Padilla-Gay}}, \ and\ \bibinfo {author} {\bibfnamefont {Irene}\ \bibnamefont
  {Tamborra}},\ }\bibfield  {title} {\enquote {\bibinfo {title} {{Neutrino
  propagation hinders fast pairwise flavor conversions}},}\ }\href {\doibase
  10.1088/1475-7516/2020/06/048} {\bibfield  {journal} {\bibinfo  {journal}
  {JCAP}\ }\textbf {\bibinfo {volume} {06}},\ \bibinfo {pages} {048} (\bibinfo
  {year} {2020})},\ \Eprint {http://arxiv.org/abs/1911.09110} {arXiv:1911.09110
  [astro-ph.HE]} \BibitemShut {NoStop}%
\bibitem [{\citenamefont {Martin}\ \emph {et~al.}(2019)\citenamefont {Martin},
  \citenamefont {Abbar},\ and\ \citenamefont {Duan}}]{Martin:2019kgi}%
  \BibitemOpen
  \bibfield  {author} {\bibinfo {author} {\bibfnamefont {Joshua~D.}\
  \bibnamefont {Martin}}, \bibinfo {author} {\bibfnamefont {Sajad}\
  \bibnamefont {Abbar}}, \ and\ \bibinfo {author} {\bibfnamefont {Huaiyu}\
  \bibnamefont {Duan}},\ }\bibfield  {title} {\enquote {\bibinfo {title}
  {{Nonlinear flavor development of a two-dimensional neutrino gas}},}\ }\href
  {\doibase 10.1103/PhysRevD.100.023016} {\bibfield  {journal} {\bibinfo
  {journal} {Phys. Rev. D}\ }\textbf {\bibinfo {volume} {100}},\ \bibinfo
  {pages} {023016} (\bibinfo {year} {2019})},\ \Eprint
  {http://arxiv.org/abs/1904.08877} {arXiv:1904.08877 [hep-ph]} \BibitemShut
  {NoStop}%
\bibitem [{\citenamefont {Richers}\ \emph {et~al.}(2021)\citenamefont
  {Richers}, \citenamefont {Willcox},\ and\ \citenamefont
  {Ford}}]{Richers:2021xtf}%
  \BibitemOpen
  \bibfield  {author} {\bibinfo {author} {\bibfnamefont {Sherwood}\
  \bibnamefont {Richers}}, \bibinfo {author} {\bibfnamefont {Donald}\
  \bibnamefont {Willcox}}, \ and\ \bibinfo {author} {\bibfnamefont {Nicole}\
  \bibnamefont {Ford}},\ }\bibfield  {title} {\enquote {\bibinfo {title}
  {{Neutrino fast flavor instability in three dimensions}},}\ }\href {\doibase
  10.1103/PhysRevD.104.103023} {\bibfield  {journal} {\bibinfo  {journal}
  {Phys. Rev. D}\ }\textbf {\bibinfo {volume} {104}},\ \bibinfo {pages}
  {103023} (\bibinfo {year} {2021})},\ \Eprint
  {http://arxiv.org/abs/2109.08631} {arXiv:2109.08631 [astro-ph.HE]}
  \BibitemShut {NoStop}%
\bibitem [{Note1()}]{Note1}%
  \BibitemOpen
  \bibinfo {note} {Throughout this paper, we use arrows to denote vectors in
  the real space and bold fonts to denote vectors in the flavor
  space.}\BibitemShut {Stop}%
\bibitem [{\citenamefont {Duan}\ \emph
  {et~al.}(2006{\natexlab{a}})\citenamefont {Duan}, \citenamefont {Fuller},\
  and\ \citenamefont {Qian}}]{Duan:2005cp}%
  \BibitemOpen
  \bibfield  {author} {\bibinfo {author} {\bibfnamefont {Huaiyu}\ \bibnamefont
  {Duan}}, \bibinfo {author} {\bibfnamefont {George~M.}\ \bibnamefont
  {Fuller}}, \ and\ \bibinfo {author} {\bibfnamefont {Yong-Zhong}\ \bibnamefont
  {Qian}},\ }\bibfield  {title} {\enquote {\bibinfo {title} {{Collective
  neutrino flavor transformation in supernovae}},}\ }\href {\doibase
  10.1103/PhysRevD.74.123004} {\bibfield  {journal} {\bibinfo  {journal} {Phys.
  Rev. D}\ }\textbf {\bibinfo {volume} {74}},\ \bibinfo {pages} {123004}
  (\bibinfo {year} {2006}{\natexlab{a}})},\ \Eprint
  {http://arxiv.org/abs/astro-ph/0511275} {arXiv:astro-ph/0511275} \BibitemShut
  {NoStop}%
\bibitem [{\citenamefont {Duan}\ \emph
  {et~al.}(2006{\natexlab{b}})\citenamefont {Duan}, \citenamefont {Fuller},
  \citenamefont {Carlson},\ and\ \citenamefont {Qian}}]{Duan:2006an}%
  \BibitemOpen
  \bibfield  {author} {\bibinfo {author} {\bibfnamefont {Huaiyu}\ \bibnamefont
  {Duan}}, \bibinfo {author} {\bibfnamefont {George~M.}\ \bibnamefont
  {Fuller}}, \bibinfo {author} {\bibfnamefont {J}~\bibnamefont {Carlson}}, \
  and\ \bibinfo {author} {\bibfnamefont {Yong-Zhong}\ \bibnamefont {Qian}},\
  }\bibfield  {title} {\enquote {\bibinfo {title} {{Simulation of Coherent
  Non-Linear Neutrino Flavor Transformation in the Supernova Environment. 1.
  Correlated Neutrino Trajectories}},}\ }\href {\doibase
  10.1103/PhysRevD.74.105014} {\bibfield  {journal} {\bibinfo  {journal} {Phys.
  Rev. D}\ }\textbf {\bibinfo {volume} {74}},\ \bibinfo {pages} {105014}
  (\bibinfo {year} {2006}{\natexlab{b}})},\ \Eprint
  {http://arxiv.org/abs/astro-ph/0606616} {arXiv:astro-ph/0606616} \BibitemShut
  {NoStop}%
\bibitem [{\citenamefont {Martin}\ \emph {et~al.}(2020)\citenamefont {Martin},
  \citenamefont {Yi},\ and\ \citenamefont {Duan}}]{Martin:2019gxb}%
  \BibitemOpen
  \bibfield  {author} {\bibinfo {author} {\bibfnamefont {Joshua~D.}\
  \bibnamefont {Martin}}, \bibinfo {author} {\bibfnamefont {Changhao}\
  \bibnamefont {Yi}}, \ and\ \bibinfo {author} {\bibfnamefont {Huaiyu}\
  \bibnamefont {Duan}},\ }\bibfield  {title} {\enquote {\bibinfo {title}
  {{Dynamic fast flavor oscillation waves in dense neutrino gases}},}\ }\href
  {\doibase 10.1016/j.physletb.2019.135088} {\bibfield  {journal} {\bibinfo
  {journal} {Phys. Lett. B}\ }\textbf {\bibinfo {volume} {800}},\ \bibinfo
  {pages} {135088} (\bibinfo {year} {2020})},\ \Eprint
  {http://arxiv.org/abs/1909.05225} {arXiv:1909.05225 [hep-ph]} \BibitemShut
  {NoStop}%
\bibitem [{\citenamefont {Raffelt}\ and\ \citenamefont
  {Smirnov}(2007)}]{Raffelt:2007xt}%
  \BibitemOpen
  \bibfield  {author} {\bibinfo {author} {\bibfnamefont {Georg~G.}\
  \bibnamefont {Raffelt}}\ and\ \bibinfo {author} {\bibfnamefont {Alexei~Yu.}\
  \bibnamefont {Smirnov}},\ }\bibfield  {title} {\enquote {\bibinfo {title}
  {{Adiabaticity and spectral splits in collective neutrino
  transformations}},}\ }\href {\doibase 10.1103/PhysRevD.76.125008} {\bibfield
  {journal} {\bibinfo  {journal} {Phys. Rev. D}\ }\textbf {\bibinfo {volume}
  {76}},\ \bibinfo {pages} {125008} (\bibinfo {year} {2007})},\ \Eprint
  {http://arxiv.org/abs/0709.4641} {arXiv:0709.4641 [hep-ph]} \BibitemShut
  {NoStop}%
\end{thebibliography}%

\end{document}